\documentclass[prd,aps,nofootinbib,superscriptaddress]{revtex4}
\usepackage{epsfig}
\usepackage{mathtools}
\usepackage{amsmath, slashed, color}
\usepackage{color}

\newcommand{\be}[1]{
\begin{eqnarray}\label{#1}}
\newcommand{\ee}{\end{eqnarray}}

\newcommand{\kperpq}{\vec{k}_\perp^{\,2}}

\allowdisplaybreaks

\begin{document}

\title{Generalized quasi parton distributions in a diquark spectator model}

\author{Shohini Bhattacharya}
\affiliation{Department of Physics, SERC,
             Temple University, Philadelphia, PA 19122, USA}

\author{Christopher Cocuzza}
\affiliation{Department of Physics, SERC,
             Temple University, Philadelphia, PA 19122, USA}

\author{Andreas Metz}
\affiliation{Department of Physics, SERC,
             Temple University, Philadelphia, PA 19122, USA}

\begin{abstract}
Recently the concept of quasi parton distributions (quasi-PDFs) for hadrons has been proposed.
Quasi-PDFs are defined through spatial correlation functions and as such can be computed numerically using quantum chromodynamics on a four-dimensional lattice.
As the hadron momentum is increased, the quasi-PDFs converge to the corresponding standard PDFs that appear in factorization theorems for many high-energy scattering processes.
Here we investigate this new concept in the case of generalized parton distributions (GPDs) by calculating the twist-2 vector GPDs in the scalar diquark spectator model.
For infinite hadron momentum, the analytical results of the quasi-GPDs agree with those of the standard GPDs.
Our main focus is to examine how well the quasi-GPDs agree with the standard GPDs for finite hadron momenta. 
We also study the sensitivity of the results on the parameters of the model.
In general, our model calculation suggests that quasi-GPDs could be a viable tool for getting information about standard GPDs.
\end{abstract}


\date{\today}

\maketitle

\section{Introduction} 
\label{s:intro}
Parton distributions (PDFs) are key quantities characterizing the quark and gluon structure of strongly interacting particles such as the nucleon~\cite{Collins:1981uw}.
Factorization theorems in quantum chromodynamics (QCD)~\cite{Collins:1989gx} allow one to extract PDFs from experimental data for a variety of high-energy scattering processes.
On the other hand, the numerical calculation of PDFs from first principles using lattice QCD has remained difficult.
The main cause of the problem is the time-dependence in the definition of PDFs in terms of light-cone correlation functions.
This affects attempts to find the full dependence of PDFs on the momentum fraction $x$ carried by the parton.
In the past, most studies of PDFs in lattice QCD have therefore focussed on $x$-moments which are given by time-independent local operators.

Recently so-called quasi-PDFs have been suggested as a way out of this problem~\cite{Ji:2013dva, Ji:2014gla}.
Quasi-PDFs, which converge to the standard PDFs if the hadron momentum $P^3 = |\vec{P}|$ is increased, are given through spatial correlation functions that can be addressed in lattice QCD.
More specifically, it has been argued that for $P^3 \to \infty$ the infrared behaviors of quasi-PDFs and standard PDFs are identical~\cite{Ji:2013dva, Ji:2014gla}.
The ultraviolet (UV) behaviors of the two types of PDFs are different though.
But this difference can be taken care of through renormalization and a perturbative matching procedure --- see for instance Refs.~\cite{Xiong:2013bka, Stewart:2017tvs, Izubuchi:2018srq} --- so that for $P^3 \to \infty$ one exactly recovers the standard PDFs.
We also mention that other approaches to compute the $x$-dependence of PDFs in lattice QCD have been proposed, some of which are closely related to the concept of quasi-PDFs~\cite{Braun:1994jq, Detmold:2005gg, Braun:2007wv, Ma:2014jla, Chambers:2017dov, Hansen:2017mnd, Radyushkin:2017cyf, Orginos:2017kos, Ma:2017pxb, Radyushkin:2017lvu, Liang:2017mye}.

In the meantime the behavior of the quasi-PDFs under renormalization~\cite{Ji:2015jwa, Ishikawa:2016znu, Chen:2016fxx, Constantinou:2017sej, Alexandrou:2017huk, Chen:2017mzz, Ji:2017oey, Ishikawa:2017faj, Green:2017xeu, Spanoudes:2018zya} and a number of further aspects of quasi-PDFs and related Euclidean correlation functions have been explored~\cite{Ji:2014hxa, Li:2016amo, Monahan:2016bvm, Radyushkin:2016hsy, Radyushkin:2017ffo, Carlson:2017gpk, Briceno:2017cpo, Xiong:2017jtn, Rossi:2017muf, Ji:2017rah, Wang:2017qyg, Chen:2017mie, Monahan:2017hpu, Radyushkin:2018cvn, Zhang:2018ggy, Ji:2018hvs, Xu:2018mpf, Jia:2018qee, Briceno:2018lfj, Rossi:2018zkn, Radyushkin:2018nbf, Ji:2018waw, Karpie:2018zaz}.
In particular, the first encouraging lattice results for quasi-PDFs and similar quantities became available~\cite{Lin:2014zya, Alexandrou:2015rja, Chen:2016utp, Alexandrou:2016jqi, Zhang:2017bzy, Alexandrou:2017huk, Chen:2017mzz, Green:2017xeu, Lin:2017ani, Orginos:2017kos, Bali:2017gfr, Alexandrou:2017dzj, Chen:2017gck, Alexandrou:2018pbm, Chen:2018xof, Chen:2018fwa, Alexandrou:2018eet, Liu:2018uuj, Bali:2018spj, Lin:2018qky}.
Moreover, several model calculations of quasi-PDFs have been carried out~\cite{Gamberg:2014zwa, Bacchetta:2016zjm, Nam:2017gzm, Broniowski:2017wbr, Hobbs:2017xtq, Broniowski:2017gfp, Xu:2018eii}.
These studies have largely focused on how well, for a given model, the quasi-PDFs describe the corresponding standard GPDs as functions of $P^3$.

In the first paper on quasi-PDFs it has already been suggested that generalized parton distributions (GPDs) could also be studied by means of spatial correlation functions (quasi-GPDs)~\cite{Ji:2013dva}. 
Over the past two decades, GPDs~\cite{Mueller:1998fv, Ji:1996ek, Radyushkin:1996nd, Ji:1996nm, Radyushkin:1996ru} have attracted enormous interest --- see Refs.~\cite{Goeke:2001tz, Diehl:2003ny, Belitsky:2005qn, Boffi:2007yc, Guidal:2013rya, Mueller:2014hsa, Kumericki:2016ehc} for reviews.
Leading-twist (twist-2) GPDs allow one to access the angular momentum of partons inside hadrons~\cite{Ji:1996ek}, and they contain information on the three-dimensional structure of hadrons~\cite{Burkardt:2000za,Ralston:2001xs,Diehl:2002he,Burkardt:2002hr}.
While GPDs in principle can be measured via hard exclusive processes such as deeply-virtual Compton scattering and meson production~\cite{Ji:1996ek, Radyushkin:1996nd, Ji:1996nm, Radyushkin:1996ru, Collins:1996fb, Collins:1998be}, it is challenging to map them out fully through experimental data.
It would therefore be useful to have further input/constraints on GPDs from lattice QCD.

Previous papers on quasi-GPDs have dealt with perturbative matching calculations~\cite{Ji:2015qla, Xiong:2015nua}. 
In the present work we study the twist-2 vector GPDs --- commonly denoted by $H$ and $E$ --- in the scalar diquark model (SDM).
In particular, we investigate how well in this approach the quasi-GPDs converge to the standard GPDs if $P^3$ is increased.

We organize the paper as follows:
In Sec.~\ref{s:GPD} we provide some kinematical relations and the definitions of quasi-GPDs.
Throughout this work we will explore two definitions of these objects.
In Sec.~\ref{s:results_analytical} we discuss the analytical results for the quasi-GPDs in the SDM and consider their forward limit.
For $P^3 \to \infty$ all analytical results of the quasi distributions reduce to those of the corresponding standard distributions.
In Sec.~\ref{s:results_numerical} we present the numerical results for the quasi distributions.
The impact of varying model parameters is explored as well.
In the large-$x$ region, for the quasi-GPDs $H_{\rm Q}$ and $E_{\rm Q}$ large hadron momenta are needed to recover the standard GPDs $H$ and $E$, respectively.
If the skewness variable is large, the quasi-GPDs converge quite well to the standard GPDs for a considerable fraction of the ERBL region.
Finally, in Sec.~\ref{s:summary} we summarize our work.

\section{GPDs: definition and kinematics}
\label{s:GPD}
We first recall the definition of the standard twist-2 vector GPDs of quarks for a spin-$\frac{1}{2}$ hadron. 
Standard GPDs are defined through the light-cone correlation function (see for instance Ref.~\cite{Diehl:2003ny})\footnote{For a generic four-vector $v$ we denote the Minkowski components by $(v^0, v^1, v^2, v^3)$ and the light-cone components by $(v^+, v^-, \vec{v}_\perp)$, with $v^+ = \frac{1}{\sqrt{2}} (v^0 + v^3)$, $v^- = \frac{1}{\sqrt{2}} (v^0 - v^3)$ and $\vec{v}_\perp = (v^1, v^2)$.} 
\begin{equation}
F^{[\Gamma]}(x, \Delta) = \frac{1}{2} \int \frac{dz^-}{2\pi} \, e^{i k \cdot z} \, \langle p' | \bar{\psi}(- \tfrac{z}{2}) \, \Gamma \, {\cal W}(- \tfrac{z}{2}, \tfrac{z}{2}) \,\psi(\tfrac{z}{2})  | p \rangle \Big|_{z^+ = 0, \vec{z}_\perp = \vec{0}_\perp} \,,
\label{e:corr_standard_GPD}
\end{equation}
with $\Gamma$ denoting a generic gamma matrix. 
The color gauge invariance of this quark-quark correlator is ensured by the Wilson line
\begin{equation}
{\cal W}(- \tfrac{z}{2}, \tfrac{z}{2}) \Big|_{z^+ = 0, \vec{z}_\perp = \vec{0}_\perp}
= {\cal P} \, \textrm{exp} \, \Bigg( - i g_s \int_{-\tfrac{z^-}{2}}^{\tfrac{z^-}{2}} \, dy^- \, A^+(0^+, y^-, \vec{0}_\perp) \Bigg) \,,
\end{equation}
where ${\cal P}$ indicates path-ordering, $g_s$ the strong coupling constant and $A^+$ the plus-component of the gluon field.
The four-momentum of the initial-state (final-state) hadron is $p$ ($p'$).
(Throughout this work we omit spin labels for the hadron.)
We also use the common kinematical variables
\begin{equation}
P = \frac{1}{2} (p + p') \,, \qquad
\Delta = p' - p \,, \qquad
t = \Delta^2 \,, \qquad
\xi = \frac{p^{\prime +} - p^+}{p^{\prime +} + p^+} = - \frac{\Delta^+}{2 P^+} \,,
\end{equation}
where for the skewness variable $\xi$ the range $0 \le \xi \le 1$ is considered.
We work in the so-called symmetric frame for which $\vec{P}_\perp = 0$, with $P^3 > 0$ large.
The Mandelstam variable $t$ can be expressed through $\xi$ and $\vec{\Delta}_\perp$,
\begin{equation}
t = - \frac{1}{1 - \xi^2} \, (4 \xi^2 M^2 + \vec{\Delta}_\perp^2) \,,
\end{equation}
with $M$ the hadron mass.
For the numerical results, $\xi$ and $\vec{\Delta}_\perp^2$ are used as independent variables.
For later convenience we also introduce the quantity
\begin{equation}
\delta = \sqrt{1 + \frac{M^2 - t/4}{(P^3)^2}} \,.
\end{equation}
By means of $P^2 = M^2 - t/4$ one then readily finds $P^0 = \delta P^3$.
For $\Gamma = \gamma^+$, Eq.~\eqref{e:corr_standard_GPD} defines the twist-2 vector GPDs $H$ and $E$,
\begin{equation}
F^{[\gamma^+]}(x, \Delta) = \frac{1}{2 P^+} \, \bar{u}(p') \bigg[ \, \gamma^+ \, H(x, \xi, t) + \frac{i \sigma^{+ \mu} \Delta_\mu}{2M} \, E(x, \xi, t) \bigg] u(p) \,,
\end{equation}
where $u(p)$ ($u(p')$) is the 4-component Dirac spinor for the incoming (outgoing) hadron, and $\sigma^{\mu\nu} = \frac{i}{2}(\gamma^\mu \gamma^\nu - \gamma^\nu \gamma^\mu)$.
In addition to $\xi$ and $t$, the standard GPDs depend on the (average) plus-momentum of the quarks $x = \frac{k^+}{P^+}$, and on a renormalization scale which we have omitted for brevity.
The support for standard GPDs is $-1 \le x \le 1$, where $\xi \le x \le 1$ ($-1 \le x \le -\xi$) is the DGLAP region for quarks (antiquarks), and 
$-\xi \le x \le \xi$ is the ERBL region.

We now turn to the quasi-GPDs which are defined through the spatial correlator~\cite{Ji:2013dva, Ji:2015qla}
\begin{equation}
F_{\rm Q}^{[\Gamma]}(x, \Delta; P^3) = \frac{1}{2} \int \frac{dz^3}{2\pi} \, e^{i k \cdot z} \, \langle p' | \bar{\psi}(- \tfrac{z}{2}) \, \Gamma \, {\cal W}_{\rm Q}(- \tfrac{z}{2}, \tfrac{z}{2}) \,\psi(\tfrac{z}{2})  | p \rangle \Big|_{z^0 = 0, \vec{z}_\perp = \vec{0}_\perp} \,,
\label{e:corr_quasi_GPD}
\end{equation}
with the Wilson line
\begin{equation}
{\cal W}_{\rm Q}(- \tfrac{z}{2}, \tfrac{z}{2}) \Big|_{z^0 = 0, \vec{z}_\perp = \vec{0}_\perp}
= {\cal P} \, \textrm{exp} \, \Bigg( - i g_s \int_{-\tfrac{z^3}{2}}^{\tfrac{z^3}{2}} \, dy^3 \, A^3(0, \vec{0}_\perp, y^3) \Bigg) \,.
\end{equation}
The correlation function in Eq.~\eqref{e:corr_quasi_GPD} depends on the ratio $x = \frac{k^3}{P^3}$, the momentum transfer $\Delta$, the (average) hadron momentum $P^3$ and a renormalization scale, which we have again omitted.\footnote{In the literature, frequently the same symbol ($x$) is used for $\frac{k^+}{P^+}$ and $\frac{k^3}{P^3}$, even though the two ratios are obviously different. Here we follow this convention.} 
Here the ratio $x$ can take any value, but if $x$ is outside the range $[-1, 1]$ the correlator is very small for large $P^3$. 
We consider two definitions of the quasi-GPDs,
\begin{eqnarray}
F^{[\gamma^0]}(x, \Delta; P^3) & = & \frac{1}{2 P^0} \, \bar{u}(p') \bigg[ \, \gamma^0 \, H_{{\rm Q}(0)}(x, \xi, t; P^3) + \frac{i \sigma^{0 \mu} \Delta_\mu}{2M} \, E_{{\rm Q}(0)}(x, \xi, t; P^3) \bigg] u(p) \,,
\label{e:quasi_GPD_0}
\\[0.1cm]
F^{[\gamma^3]}(x, \Delta; P^3) & = & \frac{1}{2 P^3} \, \bar{u}(p') \bigg[ \, \gamma^3 \, H_{{\rm Q}(3)}(x, \xi, t; P^3) + \frac{i \sigma^{3 \mu} \Delta_\mu}{2M} \, E_{{\rm Q}(3)}(x, \xi, t; P^3) \bigg] u(p) \,.
\label{e:quasi_GPD_3}
\end{eqnarray}
In the literature, so far only Eq.~\eqref{e:quasi_GPD_3} with the matrix $\gamma^3$ has been used.
But based on existing results for quasi-PDFs we want to explore both definitions.
While the original paper on quasi-PDFs suggested to use the matrix $\gamma^3$~\cite{Ji:2013dva} for the unpolarized quasi-PDF $f_{1,{\rm Q}}(x; P^3)$, it was later argued that the matrix $\gamma^0$ would lead to a better suppression of higher-twist contributions~\cite{Radyushkin:2016hsy}.
It was also found that $\gamma^0$ is preferred from the point of view of renormalization~\cite{Constantinou:2017sej}.
If one leaves aside these complications, one could work with any linear combination $\Gamma = a \gamma^3 + b \gamma^0$ with $a + b = 1$, in particular also $\Gamma = \gamma^+ / \sqrt{2}$.
In the matching calculations in Refs.~\cite{Ji:2015qla, Xiong:2015nua} the (negative of the) variable $\tilde{\xi}_3 = - \Delta^3 / (2 P^3)$ was used as the argument of the quasi-GPDs instead of $\xi$.
The two variables are related through $\tilde{\xi}_3 = \delta \xi$.
While they become identical in the limit $P^3 \to \infty$, their difference can be non-negligible for the finite $P^3$ values that we use for the numerics.
We finally note that with $P \cdot \Delta = 0$ one finds the relation $\Delta^0 = - 2 \xi P^3$ which we exploit for the analytical calculations.

\section{Analytical results in scalar diquark model}
\label{s:results_analytical}
The SDM for a relativistic spin-$\frac{1}{2}$ particle is specified through the Lagrange density 
\begin{equation}
{\cal L}_{\rm SDM} = \bar{\Psi} \big( i \, \slashed{\partial} - M \big) \Psi 
+ \bar{\psi} \big( i \, \slashed{\partial} - m_q \big) \psi
+ \frac{1}{2} \big( \partial_{\mu} \varphi \, \partial^{\mu} \varphi - m_s^2 \, \varphi^2 \big)
+ g \big( \bar{\Psi} \, \psi \, \varphi + \bar{\psi} \, \Psi \, \varphi \big) \,,
\label{e:SDM}
\end{equation}
with $\slashed{\partial} = \partial_\mu \gamma^\mu$.
In Eq.~\eqref{e:SDM}, $\Psi$ denotes the (fermionic) hadron field, $\psi$ the quark field, and $\varphi$ the scalar diquark field.
For the hadron to be stable the masses need to satisfy the relation $M < m_s + m_q$.
The main ingredient of the model is the hadron-quark-diquark vertex with the coupling constant $g$.
In this framework one can carry out perturbative calculations.
All the model results for PDFs discussed below are of ${\cal O}(g^2)$, which is the lowest nontrivial order.
We do not consider virtual diagrams which contribute for $x = 1$ only.
Diquark spectator models have been used frequently to study various aspects of the nucleon structure --- see for instance Refs.~\cite{Jakob:1997wg, Brodsky:2002cx, Meissner:2007rx, Gamberg:2007wm, Bacchetta:2008af}.
Often, scalar and vector diquarks have been involved simultaneously in order to obtain distributions of both up quarks and down quarks in the nucleon.
In addition, the nucleon-quark-diquark vertices have frequently been multiplied by form factors. 
By so doing one can eliminate UV divergences of parton correlation functions, and the model becomes more flexible due to additional parameters.
On the other hand, the model then no longer follows from a Lagrange density.
The first model calculation of quasi-PDFs has actually been carried out in such a type of diquark model~\cite{Gamberg:2014zwa} (see also Ref.~\cite{Bacchetta:2016zjm}).
The main findings of our study are not very sensitive to the type of the diquark.
We take the model as defined through Eq.~\eqref{e:SDM}, and we use a cutoff for the transverse quark momenta.
Below we briefly compare our calculation to Refs.~\cite{Gamberg:2014zwa, Bacchetta:2016zjm}.

\subsection{Results for quasi-GPDs}
We first discuss the results for the standard GPDs $H$ and $E$.
To ${\cal O}(g^2)$, one finds for the correlator in Eq.~\eqref{e:corr_standard_GPD}
\begin{equation}
F^{[\Gamma]}(x, \Delta) = \frac{i \, g^2}{2 (2\pi)^4} \int dk^- \, d^2\vec{k}_\perp \, \frac{\bar{u}(p') \, \Big( \slashed{k} + \frac{\slashed{\Delta}}{2} + m_q \Big) \, \Gamma \, \Big( \slashed{k} - \frac{\slashed{\Delta}}{2} + m_q \Big) \, u(p)}{D_{\rm GPD}}  \,, 
\end{equation}
with the denominator
\begin{equation}
D_{\rm GPD} = \bigg[ \Big( k + \frac{\Delta}{2} \Big)^2 - m_q^2 + i \varepsilon \bigg] \, \bigg[ \Big( k - \frac{\Delta}{2} \Big)^2 - m_q^2 + i \varepsilon \bigg] \, \big[ (P - k)^2 - m_s^2 + i \varepsilon \big] \,.
\end{equation}
Using Gordon identities and performing the $k^-$-integral with contour integration one obtains 
\begin{equation}
H(x, \xi, t) = 
\begin{dcases}
0 & \quad -1 \le x \le -\xi  \,, \\[0.1cm]
\frac{g^2 (x + \xi) (1 + \xi) (1 - \xi^2)}{4 (2 \pi)^3} \int d^2\vec{k}_\perp \, \frac{N_H}{D_1 \, D_2^{- \xi \le x \le \xi}}  & \quad -\xi \le x \le \xi  \,, \\[0.1cm]
\frac{g^2 (1 - x) (1 - \xi^2)}{2 (2 \pi)^3} \int d^2\vec{k}_\perp \, \frac{N_H}{D_1 \, D_2^{x \ge \xi}}  & \quad x \ge \xi \,,
\end{dcases}
\end{equation}
and a corresponding expression for the GPD $E$.
The numerators are given by
\begin{eqnarray}
N_H & = & \kperpq + (m_q + x M)^2 + (1 - x)^2 \, \frac{t}{4} - (1 - x) \xi t \, \frac{\vec{k}_\perp \cdot \vec{\Delta}_\perp}{\vec{\Delta}_\perp^2} \,,
\\[0.2cm]
N_E & = & 2 (1 - x) M \bigg[ \, m_q + \bigg( x + 2 \xi \, \frac{\vec{k}_\perp \cdot \vec{\Delta}_\perp}{\vec{\Delta}_\perp^2} \bigg) M \bigg] \,,
\end{eqnarray}
while the denominators are
\begin{eqnarray}
D_1 & = & (1 + \xi)^2 \kperpq + \frac{1}{4} (1 - x)^2 \vec{\Delta}_\perp^2 - (1 - x) (1 + \xi) \vec{k}_\perp \cdot \vec{\Delta}_\perp + (1 - x) (1 + \xi) m_q^2 + (x + \xi) (1 + \xi) m_s^2
\nonumber \\
& & - \, (1 - x) (x + \xi) M^2 \,, {\phantom{\frac{1}{4}}} 
\nonumber \\
D_2^{- \xi \le x \le \xi} & = & \xi (1 - \xi^2) \kperpq + \frac{1}{4} (1 - x^2) \xi \vec{\Delta}_\perp^2 + x (1 - \xi^2) \vec{k}_\perp \cdot \vec{\Delta}_\perp + \xi (1 - \xi^2) m_q^2 - \xi (x^2 - \xi^2) M^2 \,,
\nonumber \\
D_2^{x \ge \xi} & = & (1 - \xi)^2 \kperpq + \frac{1}{4} (1 - x)^2 \vec{\Delta}_\perp^2 + (1 - x) (1 - \xi) \vec{k}_\perp \cdot \vec{\Delta}_\perp + (1 - x) (1 - \xi) m_q^2 + (x - \xi) (1 - \xi) m_s^2 
\nonumber \\
& & - \, (1 - x) (x - \xi) M^2 \,. {\phantom{\frac{1}{4}}}
\end{eqnarray}
We repeat that standard GPDs vanish for $x$ outside the region $[-1, 1]$.  
In the SDM they also vanish for $-1 \le x \le -\xi$ since, to ${\cal O}(g^2)$, there cannot be an antiquark distribution for a fermion target.
The twist-2 standard GPDs are continuous for $x = \pm \, \xi$ even though they are given by different analytical expressions in the DGLAP and ERBL regions.
Note that spectator models typically lead to discontinuous higher-twist standard GPDs~\cite{Aslan:2018zzk}.
The analytical results for $H$ and $E$ can also be extracted from results for generalized transverse momentum dependent parton distributions given in Ref.~\cite{Meissner:2009ww}.
We find complete agreement with that work.

The model result for the quasi-GPD correlator in Eq.~\eqref{e:corr_quasi_GPD} reads
\begin{equation}
F_{\rm Q}^{[\Gamma]}(x, \Delta; P^3) = \frac{i \, g^2}{2 (2\pi)^4} \int dk^0 \, d^2\vec{k}_\perp \, \frac{\bar{u}(p') \, \Big( \slashed{k} + \frac{\slashed{\Delta}}{2} + m_q \Big) \, \Gamma \, \Big( \slashed{k} - \frac{\slashed{\Delta}}{2} + m_q \Big) \, u(p)}{D_{\rm GPD}} \,,
\end{equation}
from which, by means of Gordon identities, one obtains
\begin{equation}
H_{{\rm Q}(0/3)}(x, \xi, t; P^3) = \frac{i \, g^2 P^3}{(2\pi)^4} \int dk^0 \, d^2\vec{k}_\perp \, \frac{N_{H(0/3)}}{D_{\rm GPD}} \,, 
\label{e:HQ_SDM}
\end{equation}
with the numerators
\begin{eqnarray}
N_{H(0)} & = & \delta (k^0)^2 - \frac{2}{P^3} \bigg[ \, x (P^3)^2 - m_q M - x \, \frac{t}{4} - \frac{1}{2} \, \delta \xi t \, \frac{\vec{k}_\perp \cdot \vec{\Delta}_\perp}{\vec{\Delta}_\perp^2} \bigg] k^0 
\nonumber \\[0.1cm]
&& + \, \delta \bigg[ \, x^2 (P^3)^2 + \kperpq + m_q^2 + (1 - 2x) \, \frac{t}{4} - \delta \xi t \, \frac{\vec{k}_\perp \cdot \vec{\Delta}_\perp}{\vec{\Delta}_\perp^2} \bigg] \,,
\\[0.1cm]
N_{H(3)} & = & - \, (k^0)^2 + \frac{2}{\delta P^3} \bigg[ \, x \big( (P^3)^2 + M^2 \big) - \frac{t}{4} \bigg] k^0 - x^2 (P^3)^2 + \kperpq + m_q \big( m_q + 2 x M \big) + \frac{t}{4} - (1 - x) \, \frac{\xi t}{\delta} \, \frac{\vec{k}_\perp \cdot \vec{\Delta}_\perp}{\vec{\Delta}_\perp^2} \,. \qquad
\end{eqnarray}
The quasi-GPDs $E_{{\rm Q}(0/3)}$ are given by an expression analogous to Eq.~\eqref{e:HQ_SDM}, where the numerators are
\begin{eqnarray}
N_{E(0)} & = & - 2 M \delta \bigg( m_q + x M + 2 M \delta \xi \, \frac{\vec{k}_\perp \cdot \vec{\Delta}_\perp}{\vec{\Delta}_\perp^2}
\bigg) \bigg( \frac{k^0}{\delta P^3} - 1 \bigg) \,,
\\[0.1cm]
N_{E(3)} & = & 2 (1 - x) M \bigg( \frac{M}{\delta P^3} \, k^0 + m_q + 2 \, \frac{M \xi}{\delta} \, \frac{\vec{k}_\perp \cdot \vec{\Delta}_\perp}{\vec{\Delta}_\perp^2} \bigg) \,.
\end{eqnarray}
We carry out the $k^0$-integral using contour integration.
The poles for the quark propagator with momentum $(k - \frac{\Delta}{2})$, the quark propagator with momentum $(k + \frac{\Delta}{2})$, and the spectator propagator are given respectively by
\begin{eqnarray}
k_{1\pm}^0 & = & - \, \xi P^3 \pm \, \sqrt{(x + \delta \xi)^2 (P^3)^2 + \bigg( \vec{k}_\perp - \frac{\vec{\Delta}_\perp}{2} \bigg)^2 + m_q^2 - i \varepsilon} \,,
\label{e:pole_quark1}
\\[0.1cm]
k_{2\pm}^0 & = & \xi P^3 \pm \, \sqrt{(x - \delta \xi)^2 (P^3)^2 + \bigg( \vec{k}_\perp + \frac{\vec{\Delta}_\perp}{2} \bigg)^2 + m_q^2 - i \varepsilon} \,,
\label{e:pole_quark2}
\\[0.1cm]
k_{3\pm}^0 & = & \delta P^3 \pm \sqrt{(1 - x)^2 (P^3)^2 + \kperpq + m_s^2 - i \varepsilon} \,.
\label{e:pole_diquark}
\end{eqnarray}
We refrain from listing the explicit expressions after this integration.
We have verified that for $P^3 \to \infty$ all analytical results for the quasi-GPDs reduce to the analytical results for the corresponding standard GPDs.
This finding is an important cross check of the calculation, and it gives further support to quasi-GPDs as a tool to explore standard GPDs.
The quasi-GPDs in the model are nonzero for any value of $x$.
However, for large $P^3$ all contributions outside the region $[-\xi,1]$ are power-suppressed.
We also mention that the positions of the poles in Eqs.~\eqref{e:pole_quark1}--\eqref{e:pole_diquark} do not depend on $x$.
After the $k^0$-integral one therefore has the same functional form for any value of $x$.
This is in contrast to standard GPDs, where the position of the $k^-$-poles does depend on $x$ and, as a result, one ends up with different functional forms for the various regions even before performing the $k_\perp$-integral.
The quasi-GPDs are therefore continuous.
Since the positions of the $k^0$-poles for quasi-GPDs do not depend on the twist we argue that, in the SDM and similar approaches, higher-twist quasi-GPDs are continuous as well.

\subsection{Results for quasi-PDFs}
Based on the expressions for the GPDs one can readily obtain the results for the unpolarized forward PDF $f_1$ through the relations $f_1(x) = H(x,0,0)$ and $f_{1,{\rm Q}(0/3)}(x; P^3) = H_{{\rm Q}(0/3)}(x, 0, 0; P^3)$. 
Specifically, one finds
\begin{equation}
f_1(x) = \frac{g^2 (1 - x)}{2 (2\pi)^3} \int d^2 \vec{k}_\perp \, \frac{\kperpq + (m_q + x M)^2}{\big[ \kperpq + x m_s^2 + (1 - x) m_q^2 - x (1-x) M^2 \big]^2} \,,
\label{e:f1_SDM}
\end{equation}
which agrees with the result obtained previously (see, e.g., Ref.~\cite{Meissner:2007rx}).
Like for GPDs, the (model-independent) support of standard PDFs is $[-1,1]$, where quark PDFs for negative $x$ are related to antiquark PDFs for the corresponding positive $x$.
In our model calculation, $f_1(x)$ vanishes for negative $x$.
Since at $x = 0$ the expression in Eq.~\eqref{e:f1_SDM} is finite, $f_1(x)$ is discontinuous at this point in the SDM.
(Note that for higher-twist standard PDFs even delta function singularities at $x = 0$ can show up~\cite{Burkardt:1995ts, Kodaira:1998jn, Burkardt:2001iy, Efremov:2002qh, Pasquini:2018oyz}.)
The discontinuity of $f_1$ at $x = 0$ is not an artifact of the model, but rather in accordance with phenomenology.
To reach this conclusion we use the relation $f_1^q(-x) = - f_1^{\bar{q}}(x)$ and the fact that the unpolarized quark and antiquark distributions are positive and nonzero for $x \to 0$.

For the quasi-PDFs $f_{1,{\rm Q}(0/3)}$ one has
\begin{equation}
f_{1,{\rm Q}(0/3)}(x; P^3) = \frac{i \, g^2 P^3}{(2\pi)^4} \int dk^0 \, d^2\vec{k}_\perp \, 
\frac{N_{f1(0/3)}}{D_{\rm PDF}} \,, 
\label{e:f1Q_SDM}
\end{equation}
with the numerators
\begin{eqnarray}
N_{f1(0)} & = & \delta_0  (k^0)^2 - \frac{2}{P^3} \bigg( x (P^3)^2 - m_q M \bigg) k^0 + \delta_0 \big( x^2 (P^3)^2 + \kperpq + m_q^2 \big) \,,
\label{e:num_f10}
\\[0.1cm]
N_{f1(3)} & = & -(k^0)^2 + 2 \delta_0 x P^3 k^0 - x^2 (P^3)^2 + \kperpq + m_q \big( m_q + 2 x M \big) \,,
\label{e:num_f13}
\end{eqnarray}
and the denominator
\begin{equation}
D_{\rm PDF} = \big[k^2 - m_q^2 + i \varepsilon \big]^2 \, \big[ (P - k)^2 - m_s^2 + i \varepsilon \big] \,.
\label{e:den_PDF}
\end{equation}
In Eqs.~\eqref{e:num_f10} and~\eqref{e:num_f13} we have used the quantity $\delta_0 = \delta(t = 0)$.
The quasi-PDFs, like the quasi-GPDs, have support for any $x$, and they are continuous (including at the point $x = 0$) --- see also Ref.~\cite{Aslan_talk}.

We again use contour integration to perfom the $k^0$-integral in Eq.~\eqref{e:f1Q_SDM}, where the poles are given by the expressions in~\eqref{e:pole_quark1}--\eqref{e:pole_diquark} evaluated for $\xi = t = 0$.
In the forward limit one has double poles at $k_{1\pm}^0 = k_{2\pm}^0$. 
Closing the integration contour in the upper half plane gives contributions from the pole at $k_{3-}^0$ and the double pole at $k_{1-}^0 = k_{2-}^0$.
In the case of $f_{1,{\rm Q}(0)}$ the result of the $k^0$-integration reads
\begin{eqnarray}
f_{1,{\rm Q}(0)}(x; P^3) & = & - \, \frac{g^2 P^3}{(2\pi)^3} \int d^2 \vec{k}_\perp \bigg[
\frac{N_{f1(0)}(k_{3-}^0)}{(k_{3-}^0 - k_{1+}^0)^2 \, (k_{3-}^0 - k_{1-}^0)^2 \, (k_{3-}^0 - k_{3+}^0)}
\nonumber \\[0.1cm]
& & + \, \frac{N_{f1(0)}'(k_{1-}^0)}{(k_{1-}^0 - k_{1+}^0)^2 \, (k_{1-}^0 - k_{3+}^0) \, (k_{1-}^0 - k_{3-}^0)}
- \frac{2 \, N_{f1(0)}(k_{1-}^0)}{(k_{1-}^0 - k_{1+}^0)^3 \, (k_{1-}^0 - k_{3+}^0) \, (k_{1-}^0 - k_{3-}^0)}
\nonumber \\[0.1cm]
& & - \, \frac{N_{f1(0)}(k_{1-}^0)}{(k_{1-}^0 - k_{1+}^0)^2 \, (k_{1-}^0 - k_{3+}^0)^2 \, (k_{1-}^0 - k_{3-}^0)}
- \frac{N_{f1(0)}(k_{1-}^0)}{(k_{1-}^0 - k_{1+}^0)^2 \, (k_{1-}^0 - k_{3+}^0) \, (k_{1-}^0 - k_{3-}^0)^2}
\bigg] \,,
\label{e:f1Q_SDM_kpint}
\end{eqnarray}
where in one of the terms the derivative $N_{f1(0)}' = \frac{d}{d k^0} N_{f1(0)}$ enters.
For $P^3 \to \infty$ one can recover the standard PDF $f_1$ in Eq.~\eqref{e:f1_SDM} by using the expression in~\eqref{e:f1Q_SDM_kpint}.
In this limit, in the region $0 \le x \le 1$ only the first term in the square brackets of~\eqref{e:f1Q_SDM_kpint} is leading.
For $x > 1$ all terms are power-suppressed, while for $x < 0$ the first and last term are leading but the leading powers of the two terms cancel each other.

In order to compute standard PDFs (and GPDs for $\xi = 0$) in diquark spectator models one can use a cut-diagram approach with a single on-shell particle (diquark)~\cite{Jakob:1997wg, Brodsky:2002cx, Meissner:2007rx, Gamberg:2007wm, Bacchetta:2008af}.
In this framework, one inserts in the PDF operator a sum over a complete set of states between the quark fields and, for the calculation of real graphs to ${\cal O}(g^2)$, restricts this sum to a single diquark.
One can verify that this technique provides the same result one finds by computing the correlator without inserting a complete set of states right from the start and then performing the $k^-$-integration.
On the other hand, care has to be taken for quasi-PDFs.
To illustrate this point we consider as an example $f_{1,{\rm Q(0)}}$ in the cut-diagram approach.
One finds\footnote{In the cut-diagram approach the sign of the $i\varepsilon$ term in one of the quark propagators is different from Eq.~\eqref{e:den_PDF}. But the difference does not matter as the point $k^2 = m_q^2$ is not reached in this method.}
\begin{equation}
f_{1,{\rm Q(0),cut}}(x; P^3) = \frac{g^2}{2 (2\pi)^4} \int dk^0 \, d^2\vec{k}_\perp \, (2\pi) \, \delta \big((P - k)^2 - m_s^2 \big) \, \Theta(P^0 - k^0) \,
\frac{\bar{u}(P) \, (\slashed{k} + m_q) \, \gamma^0 \, (\slashed{k} + m_q) \, u(P)}{[k^2 - m_q^2 + i \varepsilon] \, [k^2 - m_q^2 - i \varepsilon]}  \,, 
\label{e:f1Q_SDM_cut}
\end{equation}
where the delta function and theta function ensure the on-shell diquark with positive energy.
Working out the numerator in Eq.~\eqref{e:f1Q_SDM_cut} and using
\begin{equation}
\delta \big((P - k)^2 - m_s^2 \big) \, \Theta(P^0 - k^0) = \frac{1}{k_{3+}^0 - k_{3-}^0} \, \delta(k^0 - k_{3-}^0)
\end{equation}
provides the result
\begin{eqnarray}
f_{1,{\rm Q(0),cut}}(x; P^3) & = & \frac{g^2 P^3}{(2\pi)^3} \int d^2 \vec{k}_\perp 
\frac{1}{k_{3+}^0 - k_{3-}^0} \, \frac{N_{f1(0)}(k^0)}{[k^2 - m_q^2 + i \varepsilon] \, [k^2 - m_q^2 - i \varepsilon]} \bigg|_{k^0 = k_{3-}^0}
\nonumber \\[0.1cm]
& = & - \, \frac{g^2 P^3}{(2\pi)^3} \int d^2 \vec{k}_\perp 
\frac{N_{f1(0)}(k_{3-}^0)}{(k_{3-}^0 - k_{1+}^0)^2 \, (k_{3-}^0 - k_{1-}^0)^2 \, (k_{3-}^0 - k_{3+}^0)} \,.
\label{e:f1Q_SDM_cut_kpint}
\end{eqnarray}
This expression exactly agrees with the first term on the r.h.s.~of~\eqref{e:f1Q_SDM_kpint}, while the other four terms are missing.
The discussion in the paragraph after Eq.~\eqref{e:f1Q_SDM_kpint} also implies that, for $P^3 \to \infty$, one can recover the standard PDF for $x \ge 0$, but not for $x < 0$, from the result in~\eqref{e:f1Q_SDM_cut_kpint}.
In the case of quasi-PDFs, the cut-diagram approach~\cite{Gamberg:2014zwa, Bacchetta:2016zjm} is therefore a purely phenomenological model that could be used for $x \ge 0$. 
Below we show a numerical comparison of the expressions in Eqs.~\eqref{e:f1Q_SDM_kpint} and~\eqref{e:f1Q_SDM_cut_kpint}.

\section{Numerical results in scalar diquark model}
\label{s:results_numerical}
We first show numerical results for the PDFs and then for the GPDs.
Throughout we use the coupling constant $g = 1$.
Our ``standard values" for the masses are $M = 0.939 \, \textrm{GeV}$, $m_s = 0.7 \, \textrm{GeV}$ and $m_q = 0.35 \, \textrm{GeV}$.
(Similar values for $m_s$ and $m_q$ have been used for the spectator model calculation of quasi-PDFs in Ref.~\cite{Gamberg:2014zwa}.)
We also study how sensitive our results are to variations of $m_s$ and $m_q$.
All the numerical results shown in this section are obtained with the cutoff $\Lambda = 1 \, \textrm{GeV}$ for the $k_\perp$-integration.
(Note that one needs such a UV regulator only in the case of $f_1$ and $H$.)
For the GPDs  we use $|\vec{\Delta}_\perp| = 0$ in all the plots.
We have also explored the ranges $1 \, \textrm{GeV} \le \Lambda \le 4 \, \textrm{GeV}$ and $0 \, \textrm{GeV }\le |\vec{\Delta}_\perp| \le 1 \, \textrm{GeV} $.
Our general conclusions are not affected by such variations.

\subsection{Results for quasi-PDFs}
\begin{figure}[t]
\begin{center}
\includegraphics[width=6.5cm]{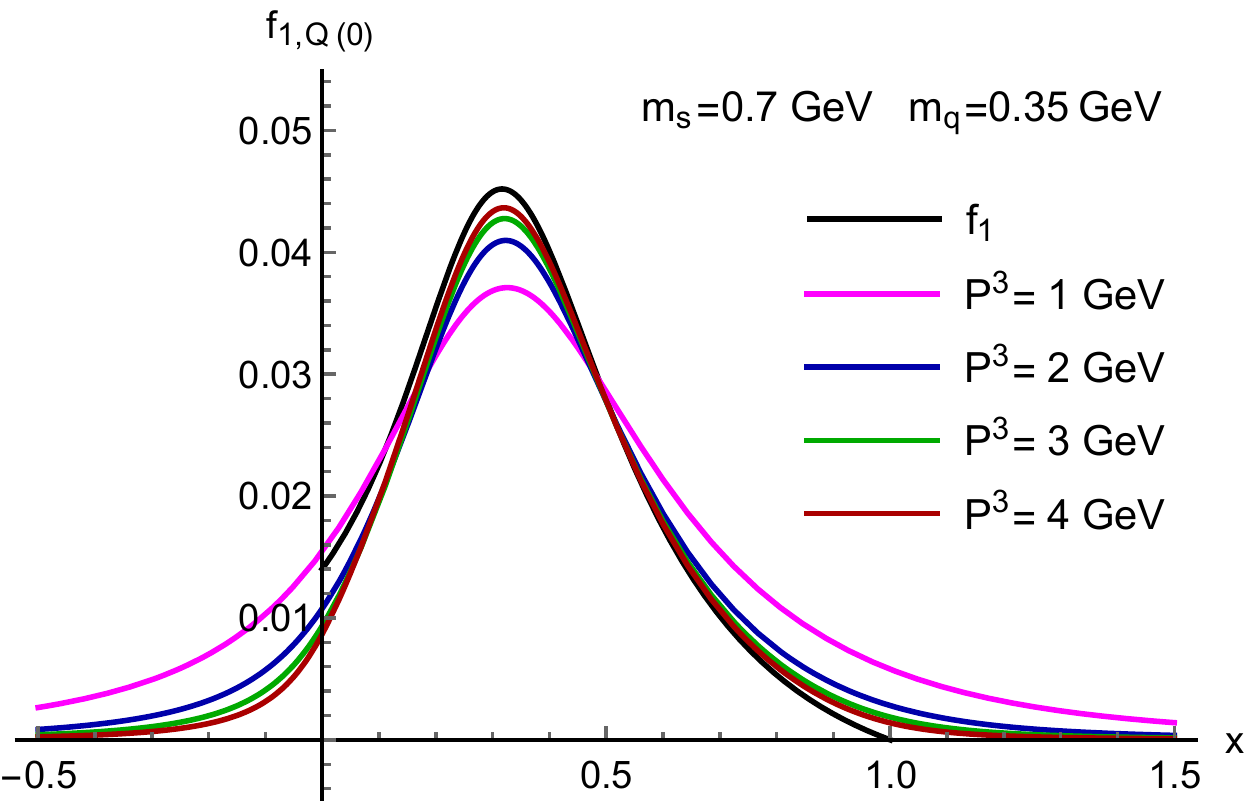}
\hspace{1.5cm}
\includegraphics[width=6.5cm]{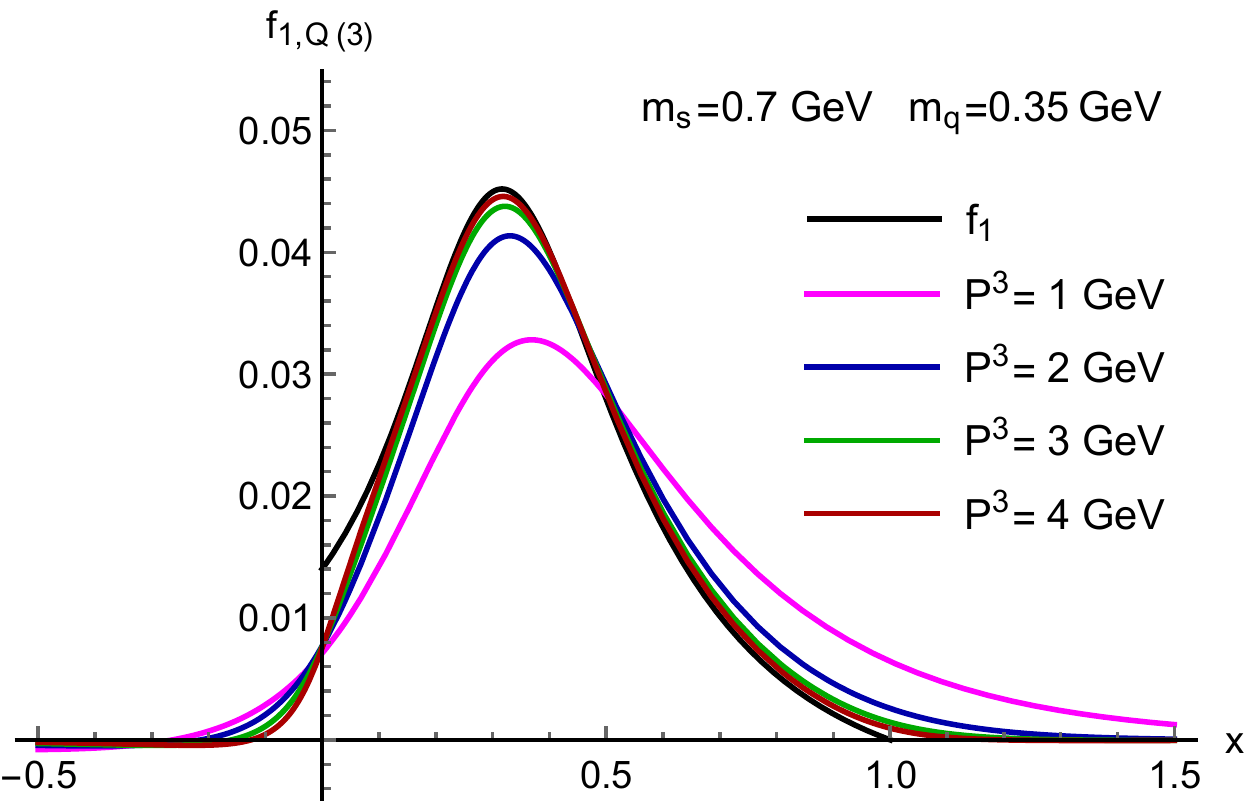}
\end{center}
\caption{Quasi-PDF $f_{1,{\rm Q}}$ as a function of $x$ for different values of $P^3$. 
Left panel: results for $f_{1,{\rm Q}(0)}$.
Right panel: results for $f_{1,{\rm Q}(3)}$.
The standard PDF $f_1$ is shown for comparison.}
\label{f:f1Q}
\end{figure}
\begin{figure}[t]
\begin{center}
\includegraphics[width=6.5cm]{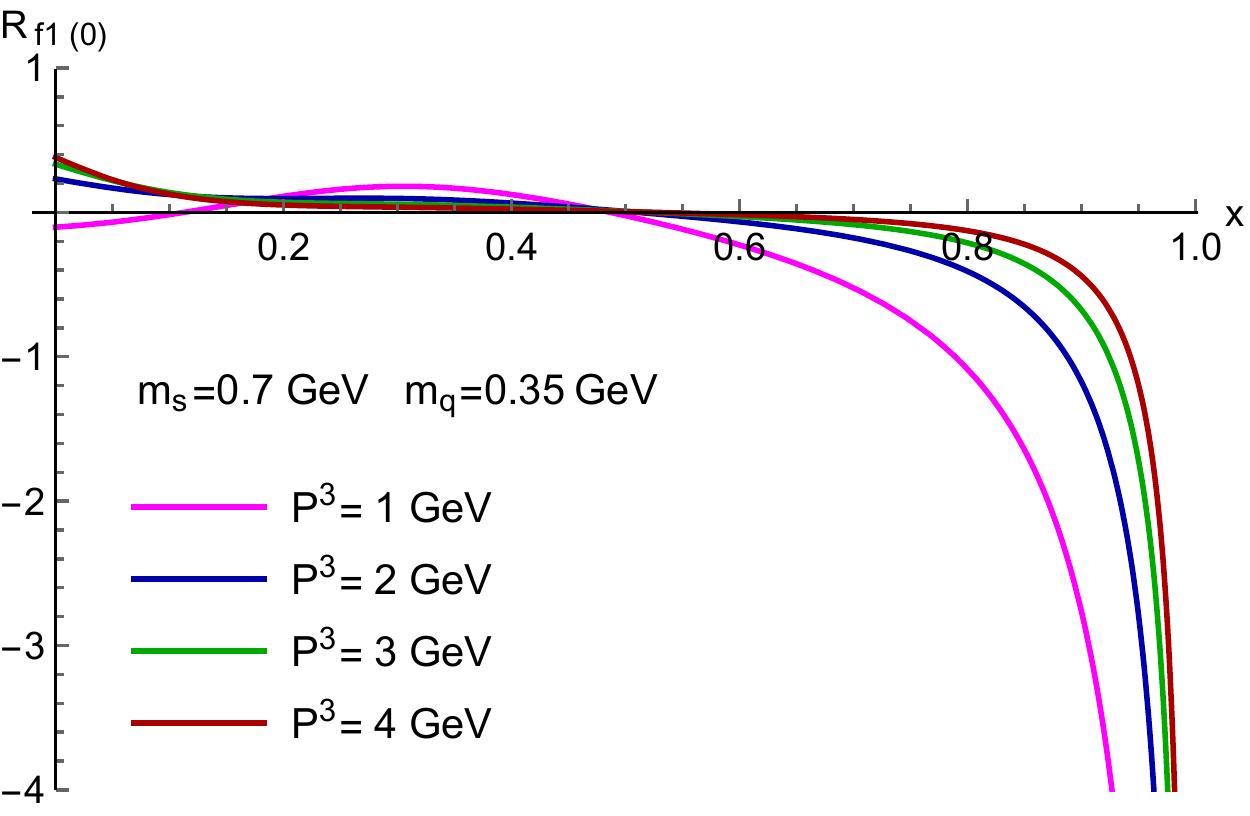}
\hspace{1.5cm}
\includegraphics[width=6.5cm]{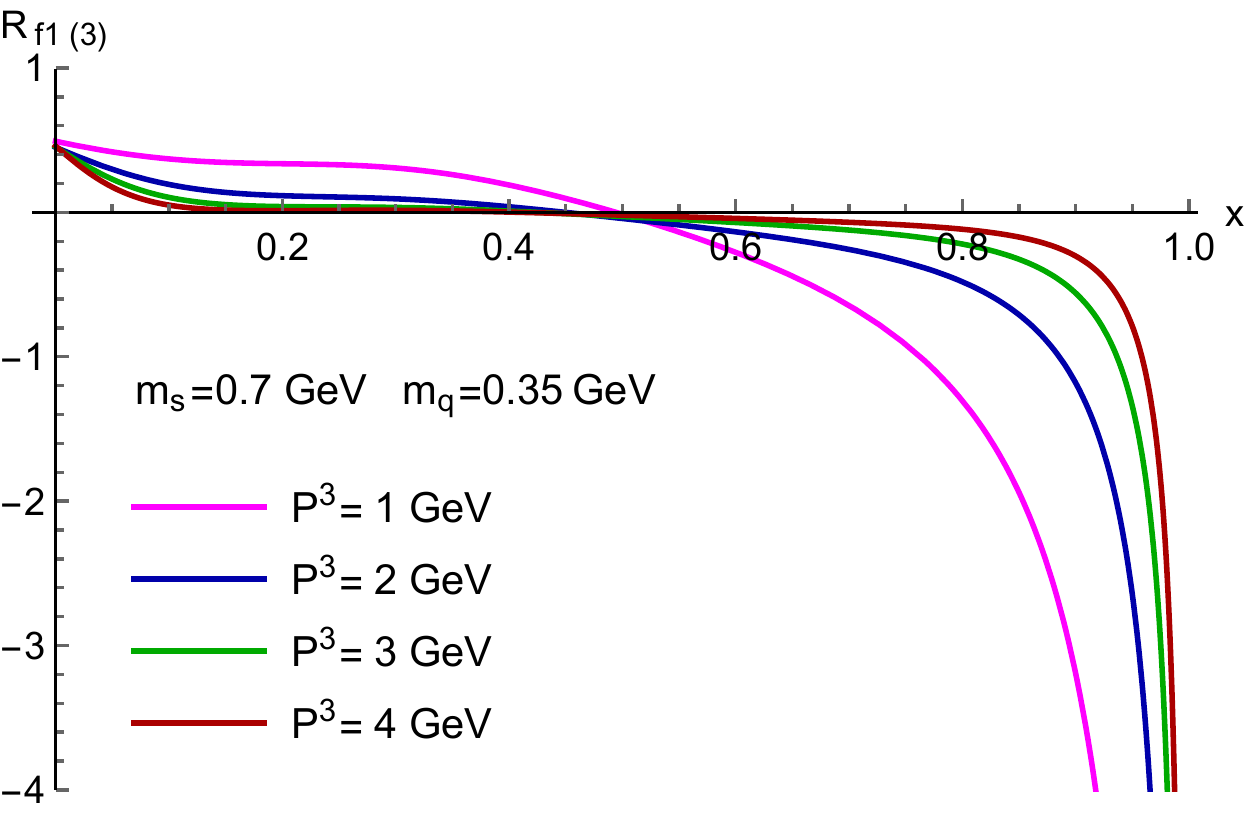}
\end{center}
\caption{Relative difference, defined in~\eqref{e:rel_difference}, between quasi-PDFs and $f_1$ as a function of $x$ for different values of $P^3$.
Left panel: results for $R_{f1(0)}$.
Right panel: results for $R_{f1(3)}$.
The maximum values of $x$ for the curves are chosen such that $|R_{f1}| \le 4$.}
\label{f:f1Q_rel}
\end{figure}
Our results for the quasi-PDFs $f_{1,{\rm Q}(0)}$ and $f_{1,{\rm Q}(3)}$ are shown in Fig.~\ref{f:f1Q} for different values of $P^3$, along with the standard PDF $f_1$.
For hadron momenta of about $2 \, \textrm{GeV}$ or larger there is not much difference between using the matrix $\gamma^0$ or $\gamma^3$.
(Note that $P^3 \approx 2 \, \textrm{GeV}$ seems within reach for current calculations of quasi-PDFs in lattice QCD~\cite{Alexandrou:2018pbm, Chen:2018xof}.)
This is actually a general outcome of our study; only for the quasi-GPD $E_{\rm Q}$ somewhat larger differences show up between the two definitions, as discussed below.
For larger $P^3$, both quasi-PDFs are quite close to $f_1$ in a wide $x$-range.
But for large $x$ striking discrepancies exist.
To better illustrate this point Fig.~\ref{f:f1Q_rel} shows the relative difference
\begin{equation}
R_{f1(0/3)}(x; P^3) = \frac{f_1(x) - f_{1,{\rm Q}(0/3)}(x; P^3)}{f_1(x)}
\label{e:rel_difference}
\end{equation}
between the quasi-PDFs and $f_1$.
According to the model, for $P^3 = 2 \, \textrm{GeV}$ one can hardly go above $x = 0.8$ if one wants to keep this difference below $50 \%$. 
For the quasi-PDFs $f_{1,{\rm Q}(3)}$ of the nucleon, the very same problem with the large-$x$ region has already been found in a different version of the diquark model~\cite{Gamberg:2014zwa} --- see also Ref.~\cite{Bacchetta:2016zjm}.
The authors of Ref.~\cite{Xu:2018eii} have also observed significant discrepancies at large $x$ for model results of quasi-PDFs of pions and kaons.
It is important that in lattice QCD such discrepancies occur as well, but they may be reduced through, in particular, the matching procedure --- see for instance Refs.~\cite{Alexandrou:2018pbm, Chen:2018xof}.
While the matching still has uncertainties, the general situation seems encouraging even for the large-$x$ region.
Of course more work is needed to investigate this crucial point.
One also finds that nonnegligible differences occur between the quasi-PDFs and $f_1$ for small $x$.
This result is not surprising since $f_1$ is discontinuous at $x = 0$, whereas the quasi-PDFs are continuous and for large $P^3$ have to approach zero for negative $x$ --- see also the paragraph after Eq.~\eqref{e:f1_SDM}.
\begin{figure}[t]
\begin{center}
\includegraphics[width=6.5cm]{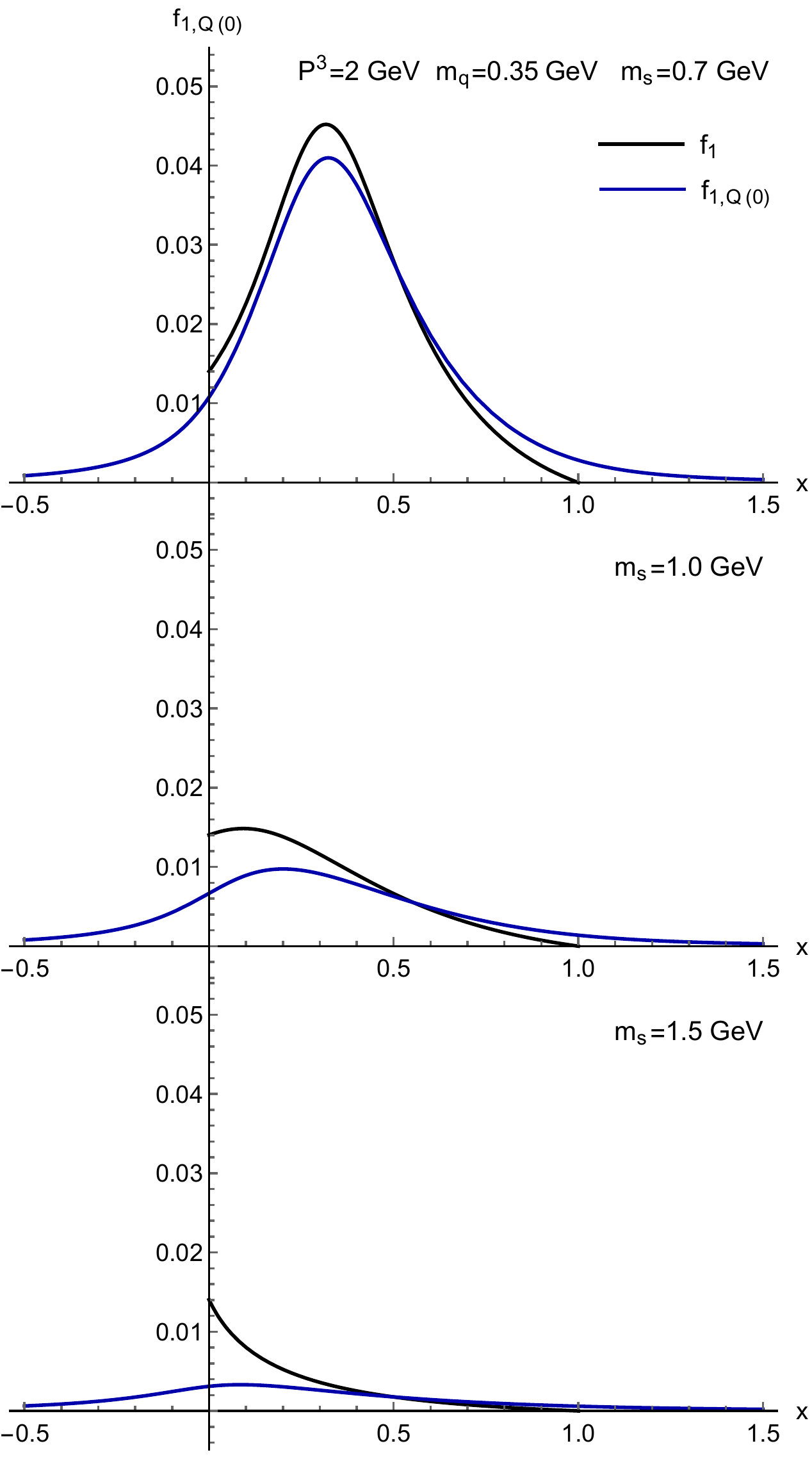}
\hspace{1.5cm}
\includegraphics[width=6.5cm]{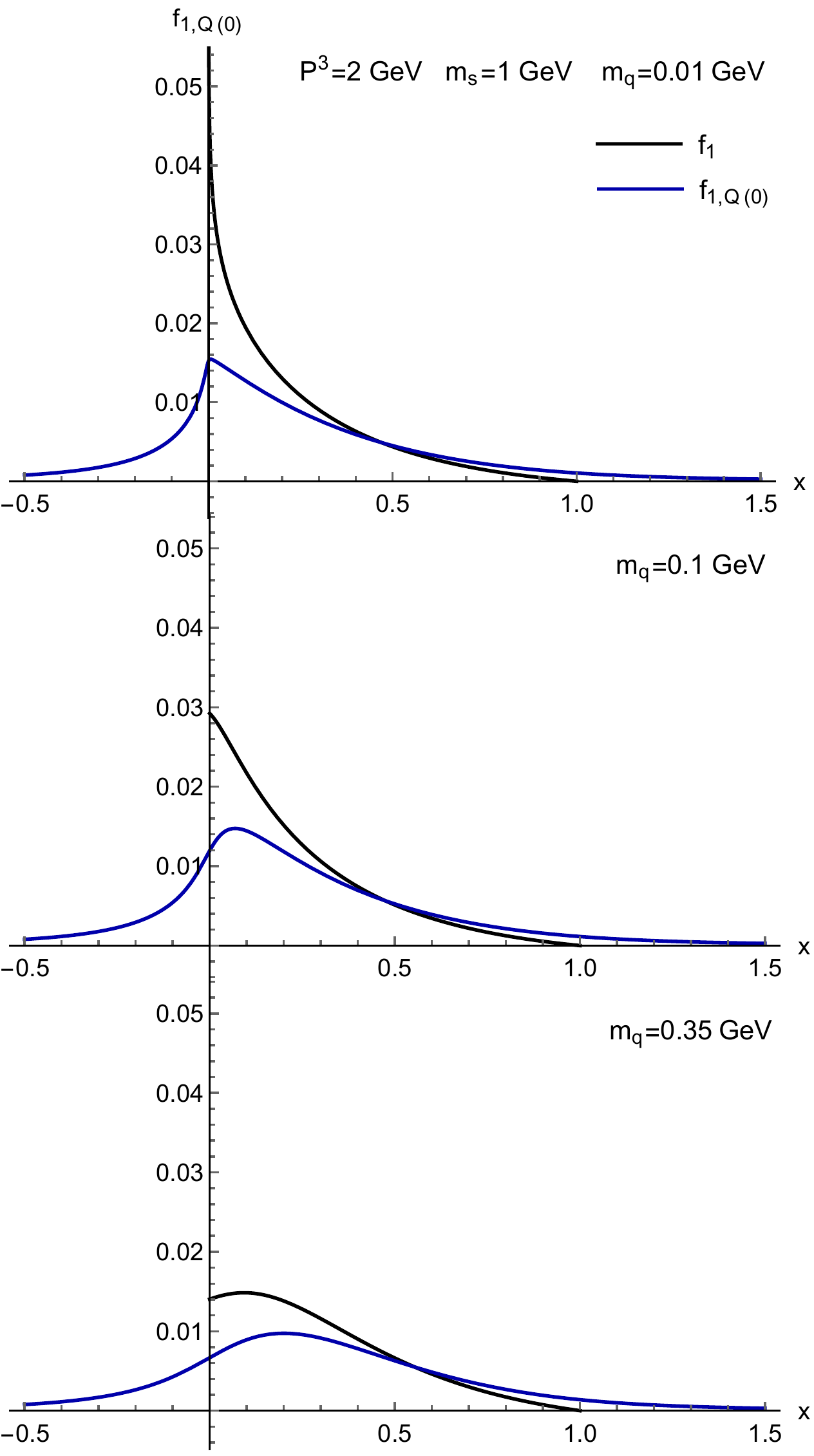}
\end{center}
\caption{Parameter dependence of quasi-PDF $f_{1,{\rm Q}(0)}$. 
Left panel: dependence on $m_s$.
Right panel: dependence on $m_q$.
All results are for $P^3 = 2 \, \textrm{GeV}$.
The standard PDF $f_1$ is shown for comparison.}
\label{f:f1Q_mass}
\end{figure}
\begin{figure}[!]
\begin{center}
\includegraphics[width=6.5cm]{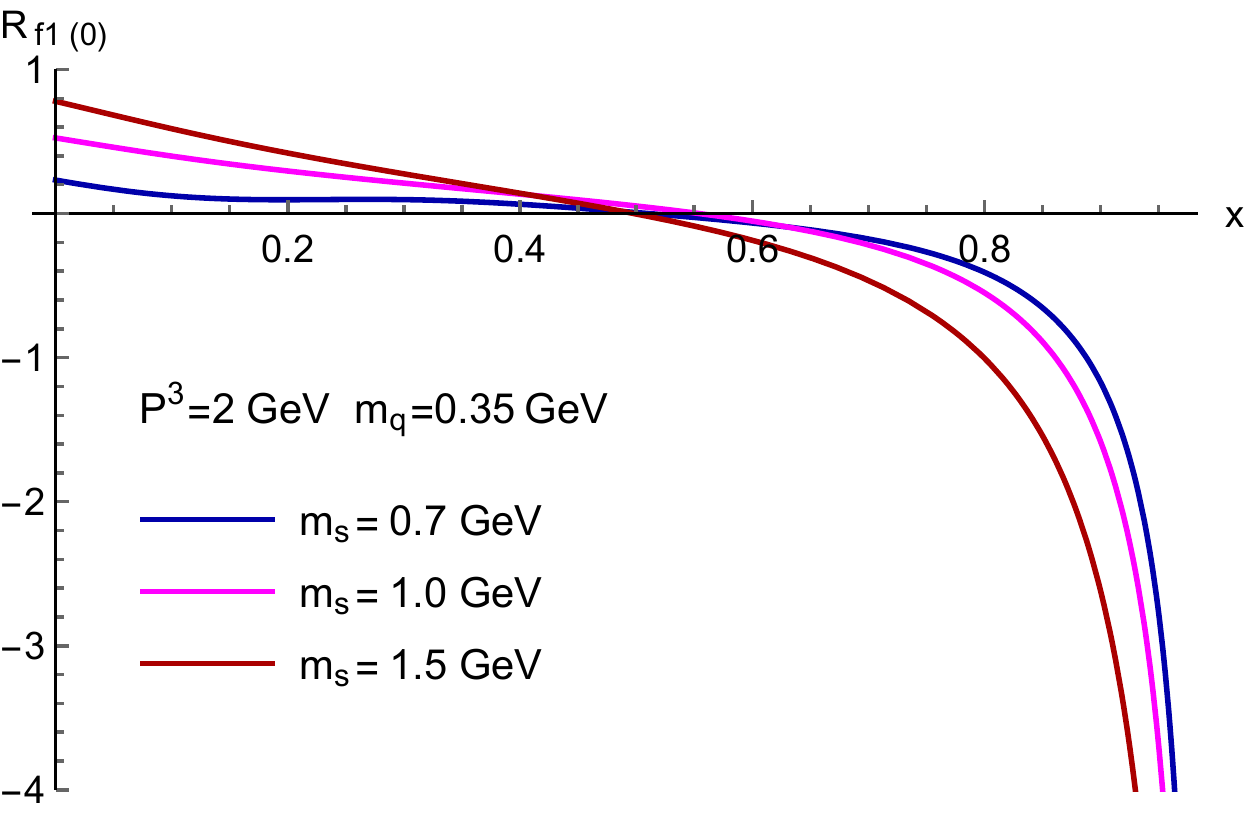}
\hspace{1.5cm}
\includegraphics[width=6.5cm]{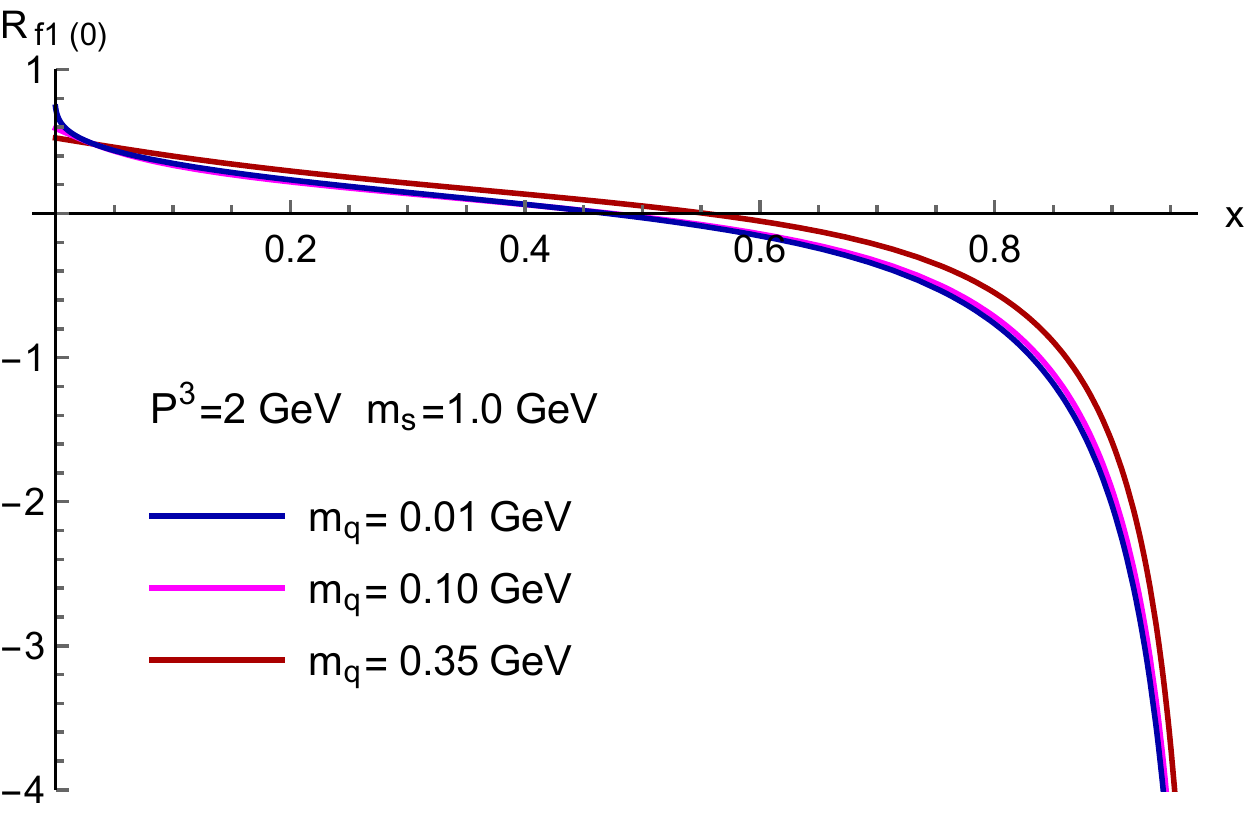}
\end{center}
\caption{Parameter dependence of relative difference between quasi-PDF $f_{1,{\rm Q}(0)}$ and $f_1$.
Left panel: dependence on $m_s$.
Right panel: dependence on $m_q$.
All results are for $P^3 = 2 \, \textrm{GeV}$.}
\label{f:f1Q_rel_mass}
\end{figure}
\begin{figure}[t]
\begin{center}
\includegraphics[width=6.5cm]{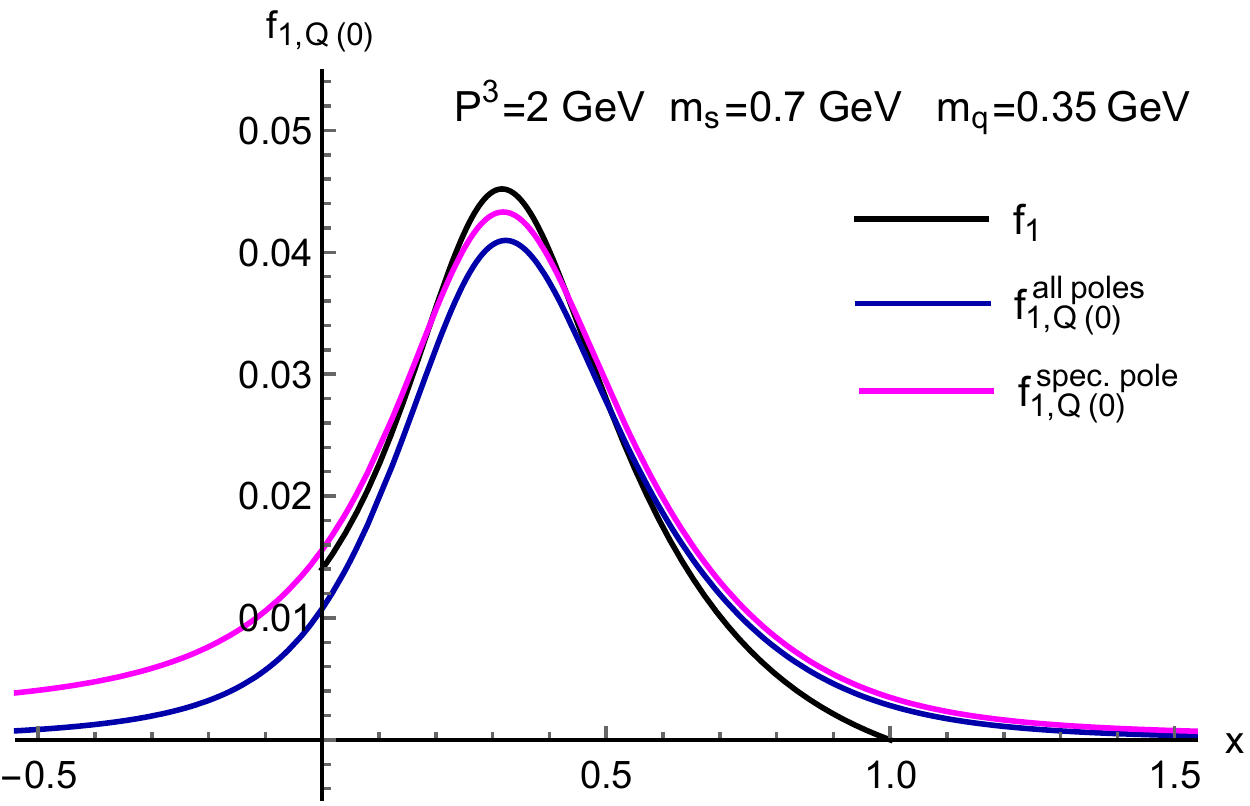}
\hspace{1.5cm}
\includegraphics[width=6.5cm]{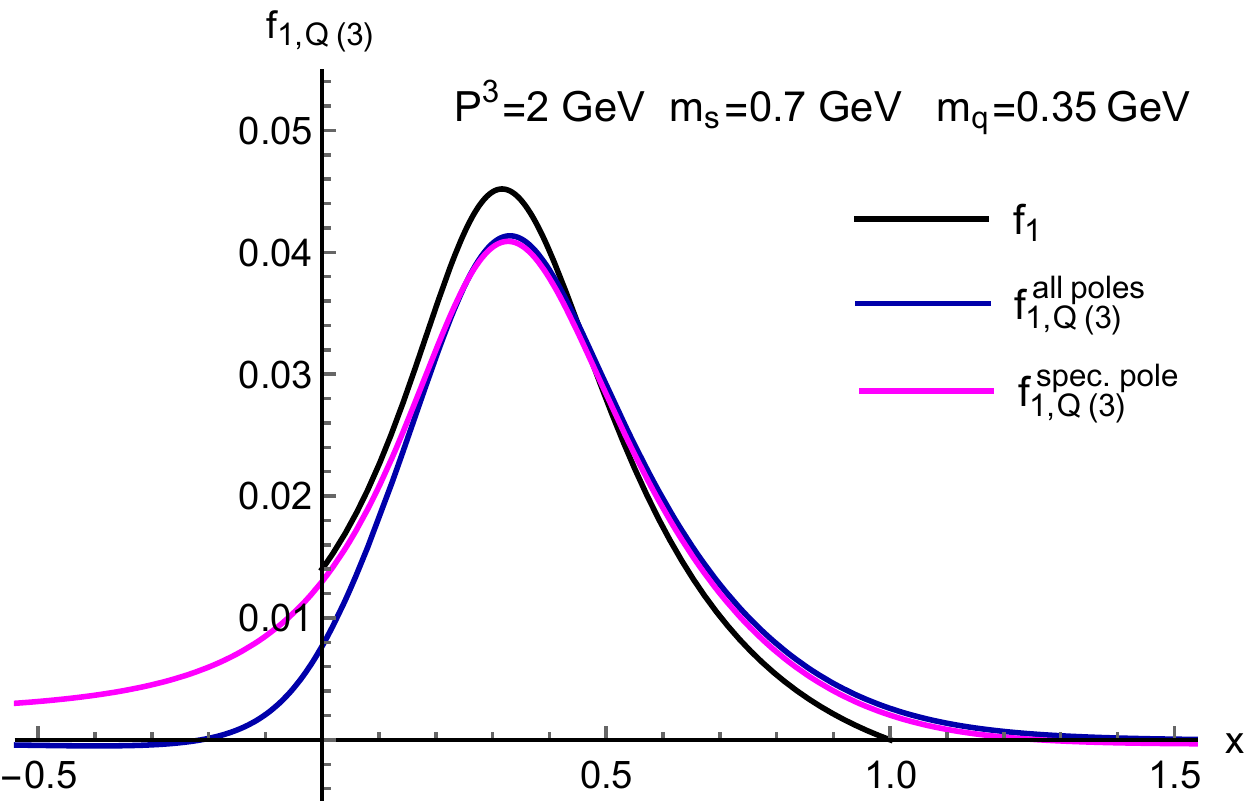}
\end{center}
\caption{Comparison between full results for the quasi-PDFs ($f_{1,{\rm Q}}^{\textrm{all poles}}$) and results from cut-diagram approach ($f_{1,{\rm Q}}^{\textrm{spec.~pole}}$).
Left panel: comparison for $f_{1,{\rm Q}(0)}$.
Right panel: comparison for $f_{1,{\rm Q}(3)}$. 
All results are for $P^3 = 2 \, \textrm{GeV}$.}
\label{f:f1Q_cut}
\end{figure}
\begin{figure}[t]
\begin{center}
\includegraphics[width=6.5cm]{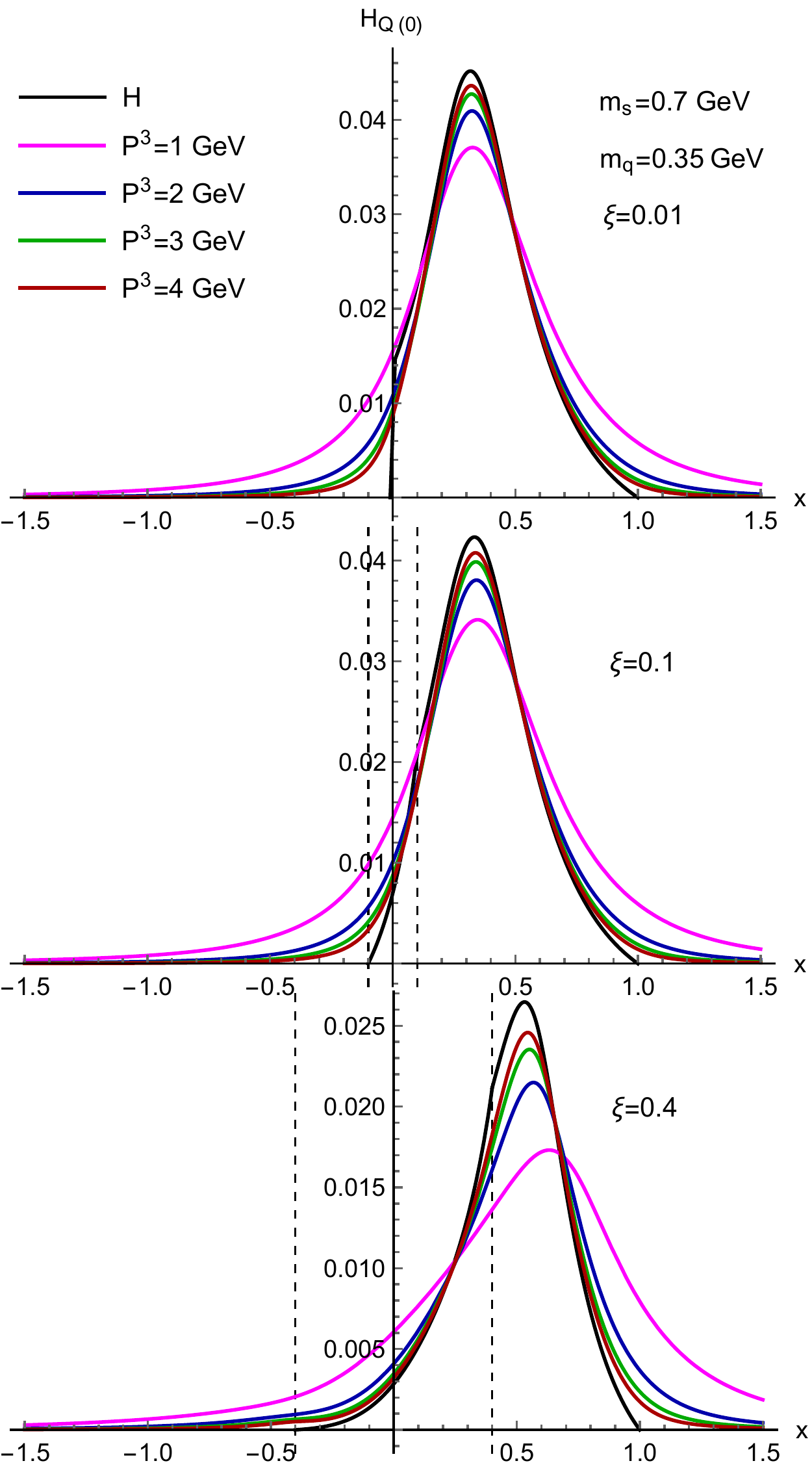}
\hspace{1.5cm}
\includegraphics[width=6.5cm]{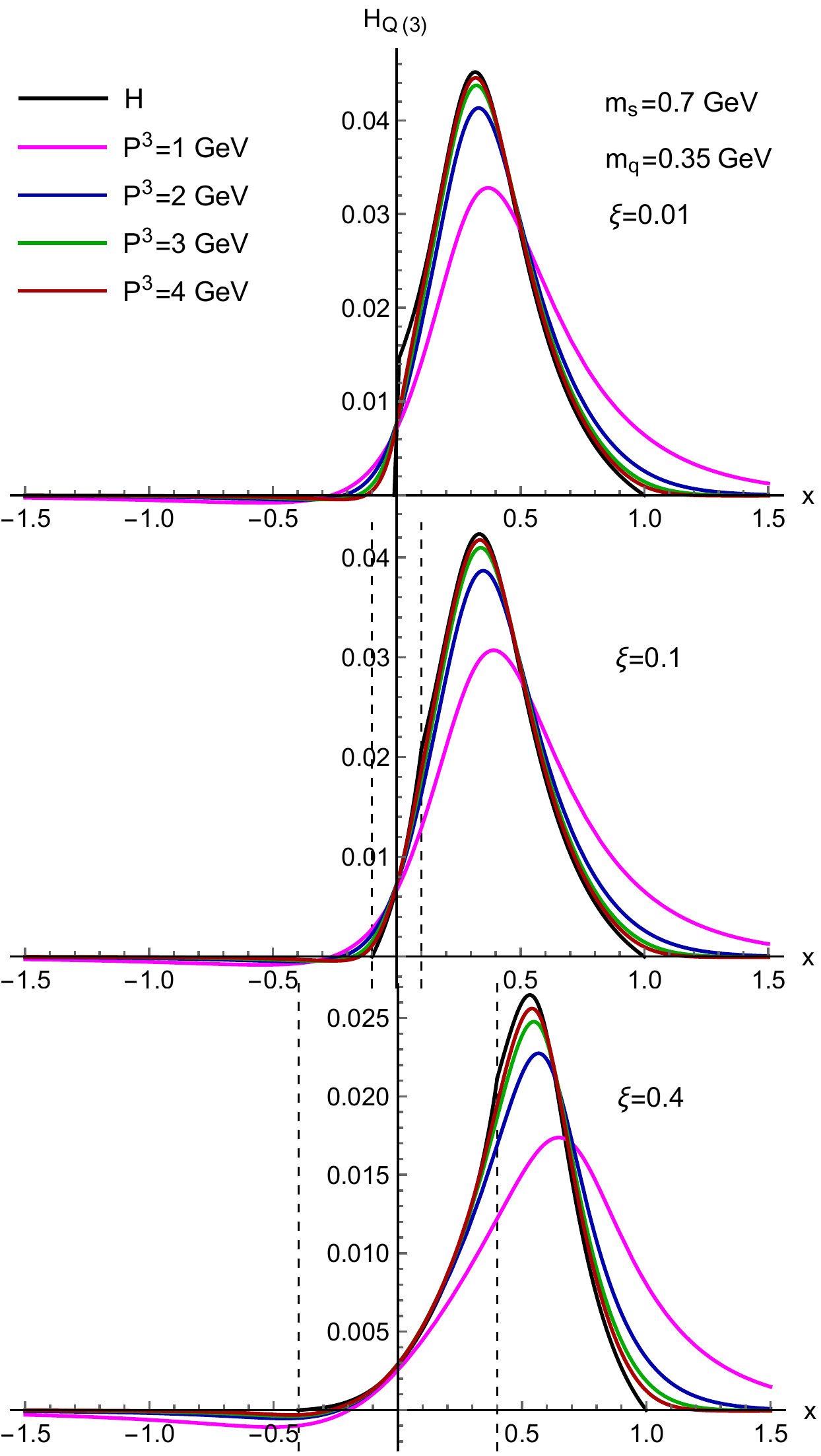}
\end{center}
\caption{Quasi-GPD $H_{\rm Q}$ as a function of $x$ for different values of $\xi$ and $P^3$. 
Left panel: results for $H_{{\rm Q}(0)}$.
Right panel: results for $H_{{\rm Q}(3)}$.
The standard GPD $H$ is shown for comparison.
The limits of the ERBL region are indicated by vertical dashed lines.}
\label{f:HQ}
\end{figure}

Information about the dependence of the results on the masses $m_s$ and $m_q$ is given in Fig.~\ref{f:f1Q_mass} and in Fig.~\ref{f:f1Q_rel_mass}.
Both $f_1$ and $f_{1,{\rm Q}(0)}$ change significantly if either $m_s$ is increased or $m_q$ is decreased.
While varying $m_s$ affects the distributions in the entire $x$-range, varying $m_q$ mostly affects the small-$x$ region only.
However, the dependence of the relative difference on changing the masses is milder.
In fact there is hardly any dependence of $R_{f1(0)}$ on $m_q$.
On the other hand, $R_{f1(0)}$ increases when $m_s$ increases.
We find the same general outcome for the quasi-PDF $f_{1,{\rm Q}(3)}$ and even for the quasi-GPDs.
One could therefore conclude that our standard values for the masses are the ``optimal choice" in order to minimize the difference between quasi and standard distributions. 

Finally, in Fig.~\ref{f:f1Q_cut} we compare the full results for the quasi-PDFs with those obtained in the cut-diagram approach --- see also the related discussion in Sec.~\ref{s:results_analytical}.
In the case of $f_{1,{\rm Q}(0)}$ the analytical expressions are listed in Eqs.~\eqref{e:f1Q_SDM_kpint} and~\eqref{e:f1Q_SDM_cut_kpint}.
Obvious modification of these equations gives the corresponding results for $f_{1,{\rm Q}(3)}$.
For positive and not too small $x$, making the approximation of keeping the spectator pole only does not have much influence.
But more deviation occurs as $x \to 0$, and the quasi-PDFs computed in the cut-diagram approach actually get closer to $f_1$.
On the other hand, this method cannot be used for $x < 0$.
We repeat that, even for large $P^3$, in the negative-$x$ region $f_{1,{\rm Q}(0/3)}^{\textrm{spec.~pole}}$ does not tend to zero.

\subsection{Results for quasi-GPDs}
\begin{figure}[t]
\begin{center}
\includegraphics[width=6.5cm]{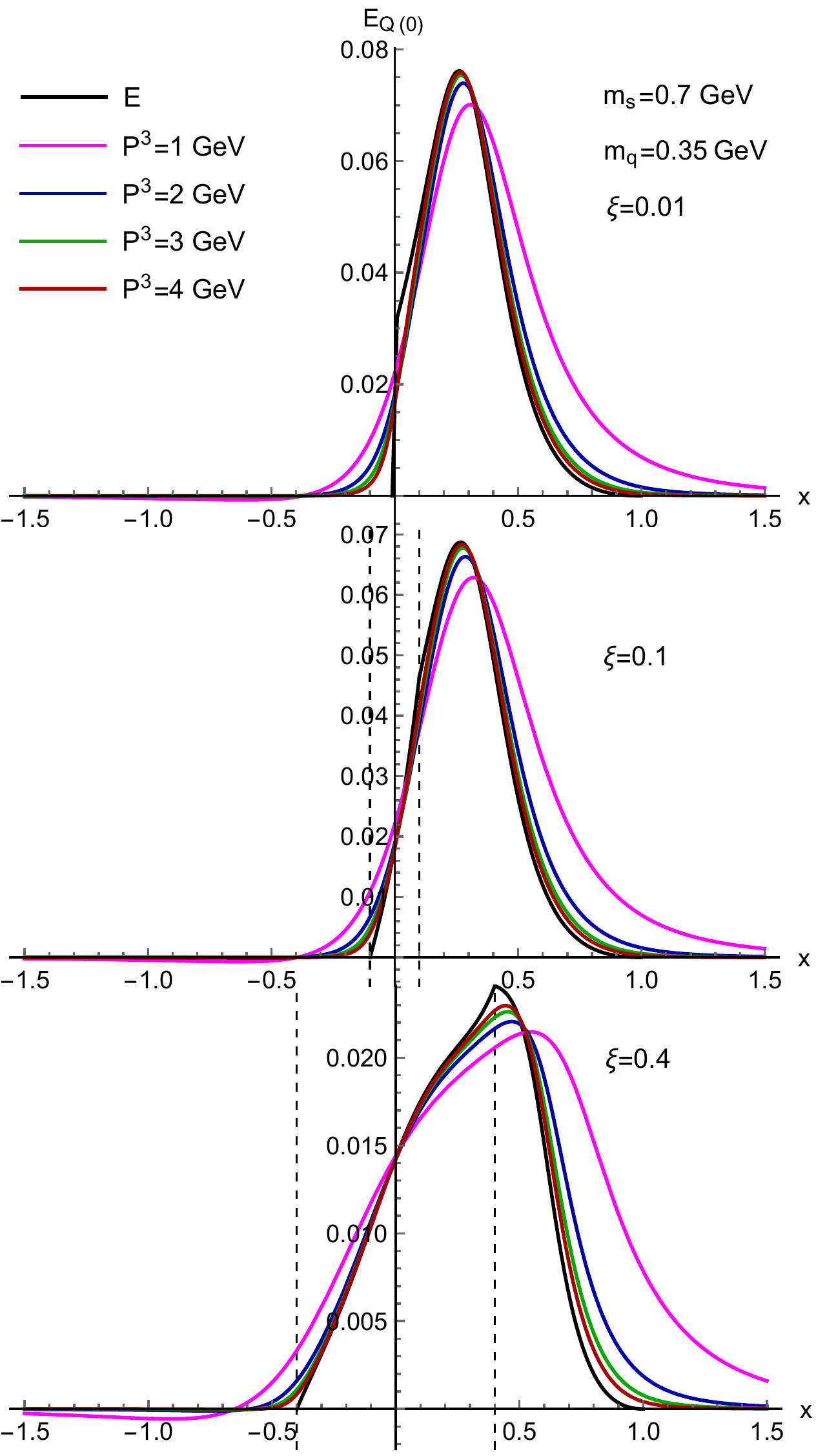}
\hspace{1.5cm}
\includegraphics[width=6.5cm]{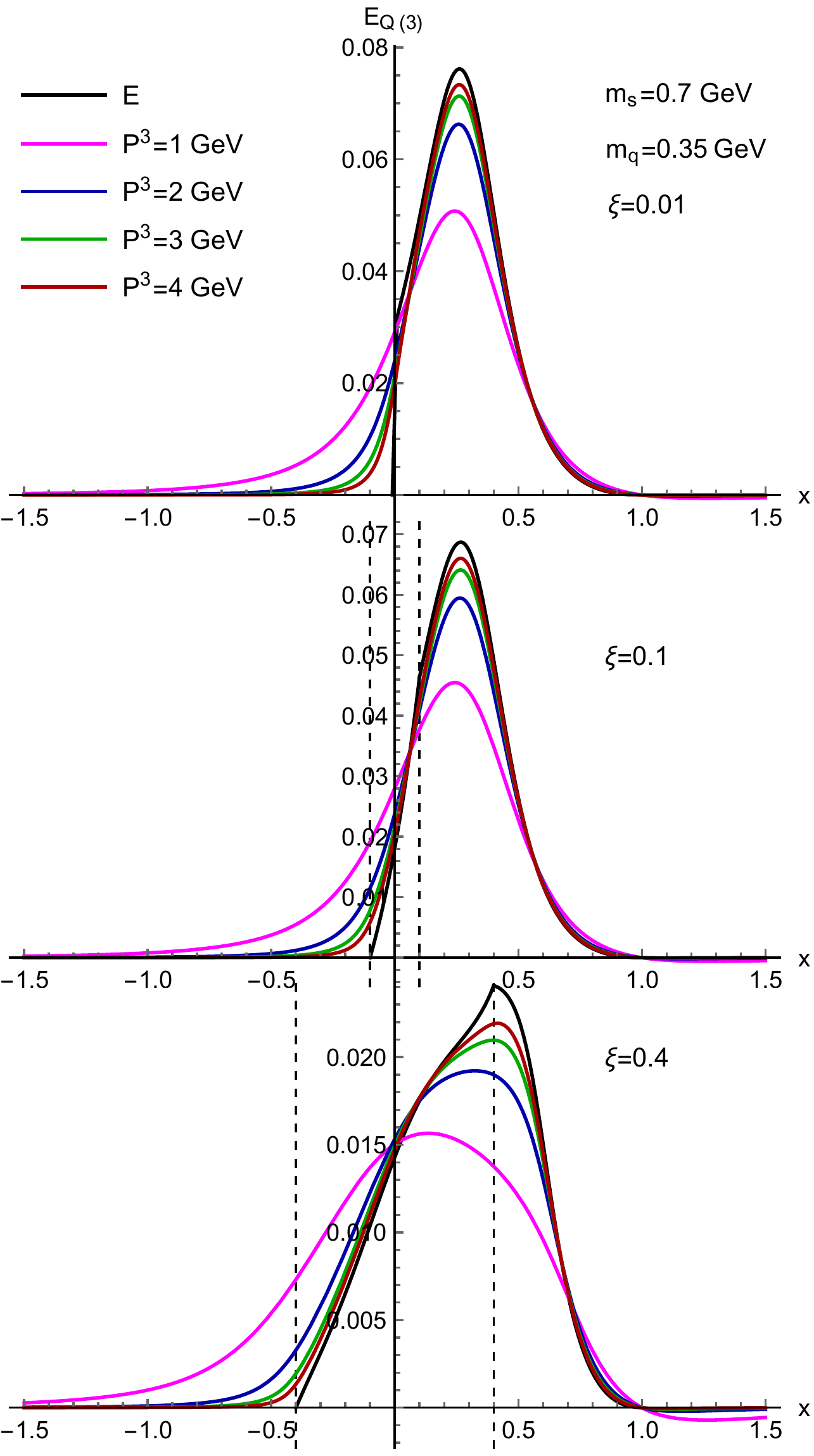}
\end{center}
\caption{Quasi-GPD $E_{\rm Q}$ as a function of $x$ for different values of $\xi$ and $P^3$. 
Left panel: results for $E_{{\rm Q}(0)}$.
Right panel: results for $E_{{\rm Q}(3)}$.
The standard GPD $E$ is shown for comparison.
The limits of the ERBL region are indicated by vertical dashed lines.}
\label{f:EQ}
\end{figure}
In Fig.~\ref{f:HQ} and Fig.~\ref{f:EQ}, the results for the quasi GPDs $H_{\rm Q}$ and $E_{\rm Q}$ are shown, respectively.
As mentioned above already, for $P^3 \gtrsim 2 \, \textrm{GeV}$ it does not matter very much whether one uses the definition involving $\gamma^0$ or $\gamma^3$.
Upon a closer look one finds that $E_{{\rm Q}(3)}$ better describes the standard GPD $E$ in the large-$x$ region, while $E_{{\rm Q}(0)}$ does better in that regard for smaller moderate $x$.
Like for the quasi-PDFs, at very large $x$ the quasi-GPDs do not converge well to the respective standard GPDs.
One could have anticipated this outcome for $H_{\rm Q}$ (due to the relation to $f_{1,{\rm Q}}$) but not necessarily for $E_{\rm Q}$.
In general, the mismatch at large $x$ grows with increasing skewness.

\begin{figure}[t]
\begin{center}
\includegraphics[width=6.5cm]{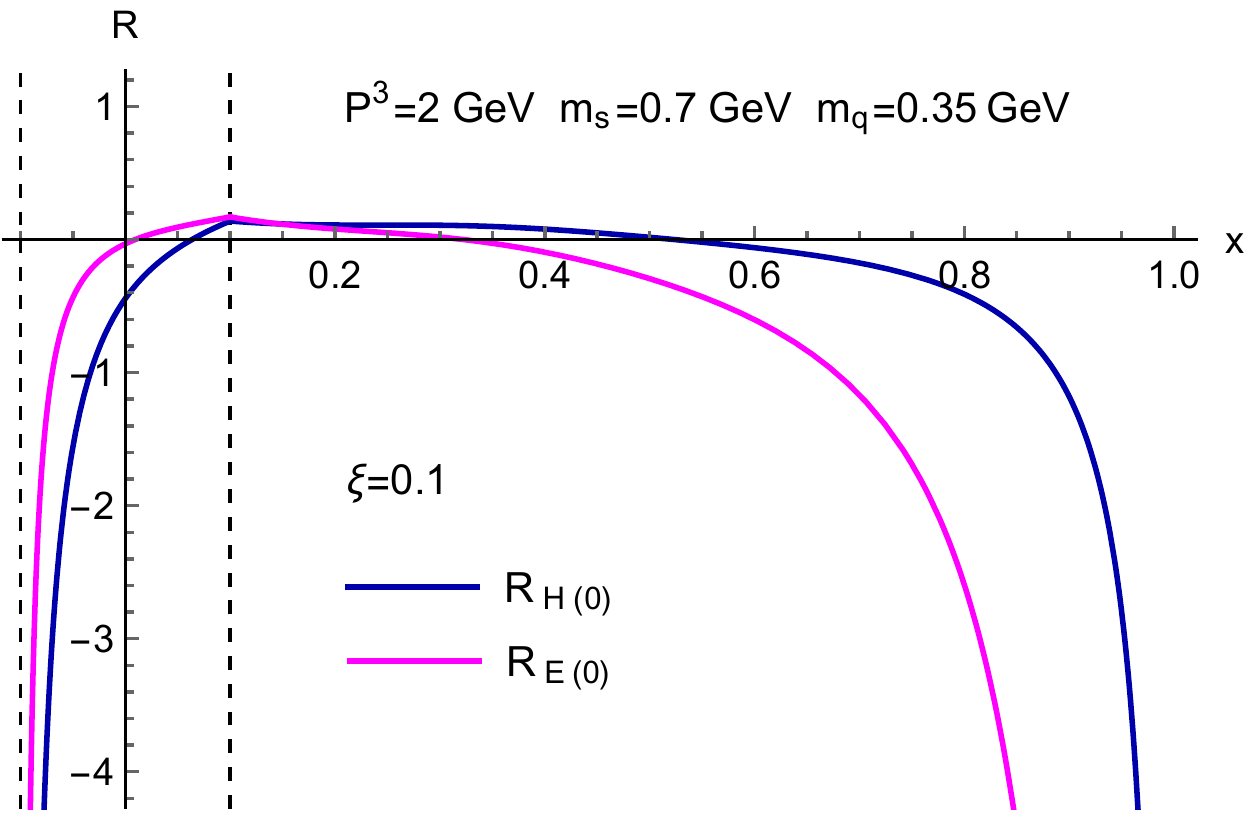}
\hspace{1.5cm}
\includegraphics[width=6.5cm]{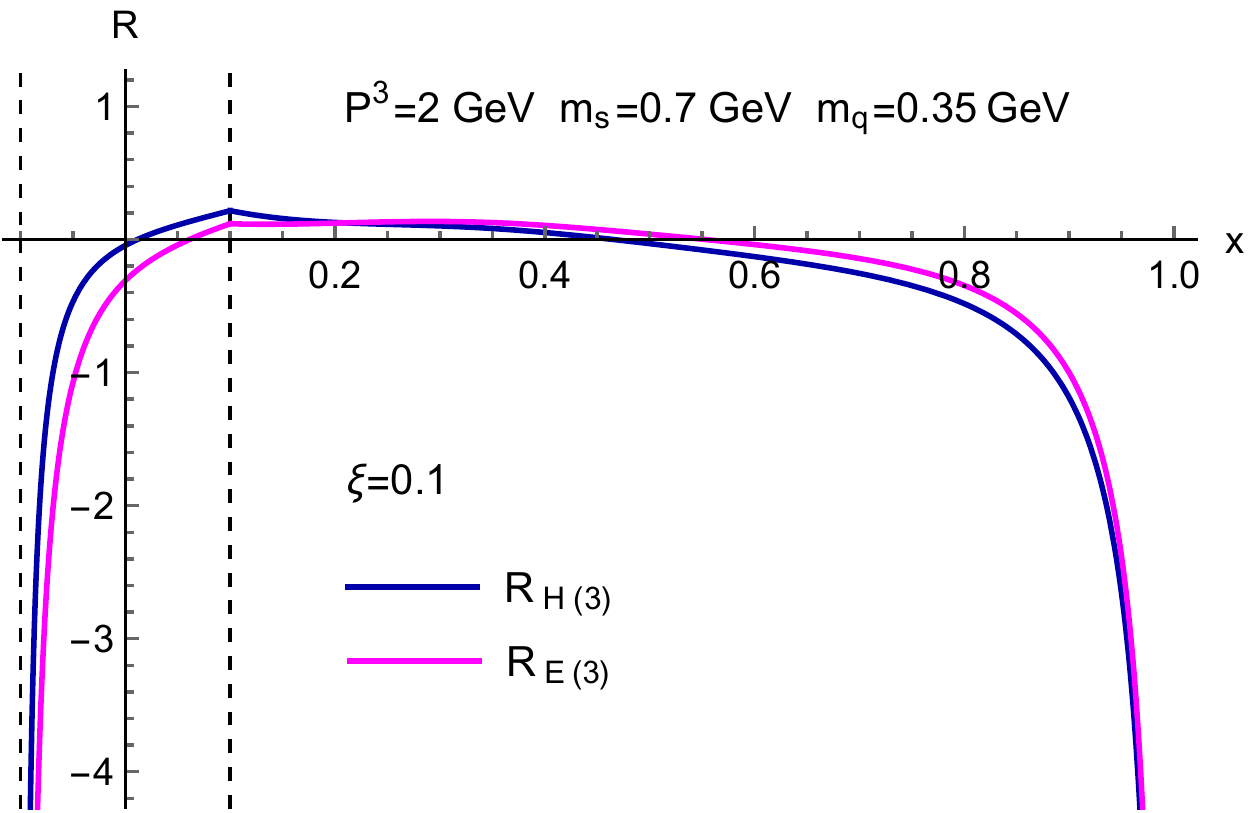}
\end{center}
\caption{Relative difference, defined analogous to Eq.~\eqref{e:rel_difference}, between quasi-GPDs and standard GPDs as a function of $x$ for $\xi = 0.1$.
Left panel: results for $R_{H(0)}$ and $R_{E(0)}$.
Right panel: results for $R_{H(3)}$ and $R_{E(3)}$.
All results are for $P^3 = 2 \, \textrm{GeV}$.}
\label{f:HEQ_rel}
\end{figure}

To better visualize, especially for large $x$, how the quasi-GPDs and standard GPDs compare we show for $\xi = 0.1$ their relative difference in Fig.~\ref{f:HEQ_rel}.
The plots in that figure also allow one to directly compare the behavior of $H_{\rm Q}$ and $E_{\rm Q}$.
At large $x$, the results for the relative difference are overall very similar to the PDF case.
Moreover, there is apparently no big difference between $H_{\rm Q}$ and $E_{\rm Q}$.
(We do not read too much into the outcome that $E_{{\rm Q}(0)}$ behaves poorer in the large-$x$ region than the other quasi-GPDs.)
We repeat that for quasi-PDFs obtained in lattice QCD the matching procedure could lead to a better description of the large-$x$ behavior. 
In Ref.~\cite{Ji:2015qla} it has been argued that for the GPD $E_{\rm Q}$ no nontrivial matching exists.
Whether it is therefore harder to find good results at large $x$ for the GPD $E$ in lattice QCD remains to be seen.

\begin{figure}[t]
\begin{center}
\includegraphics[width=6.5cm]{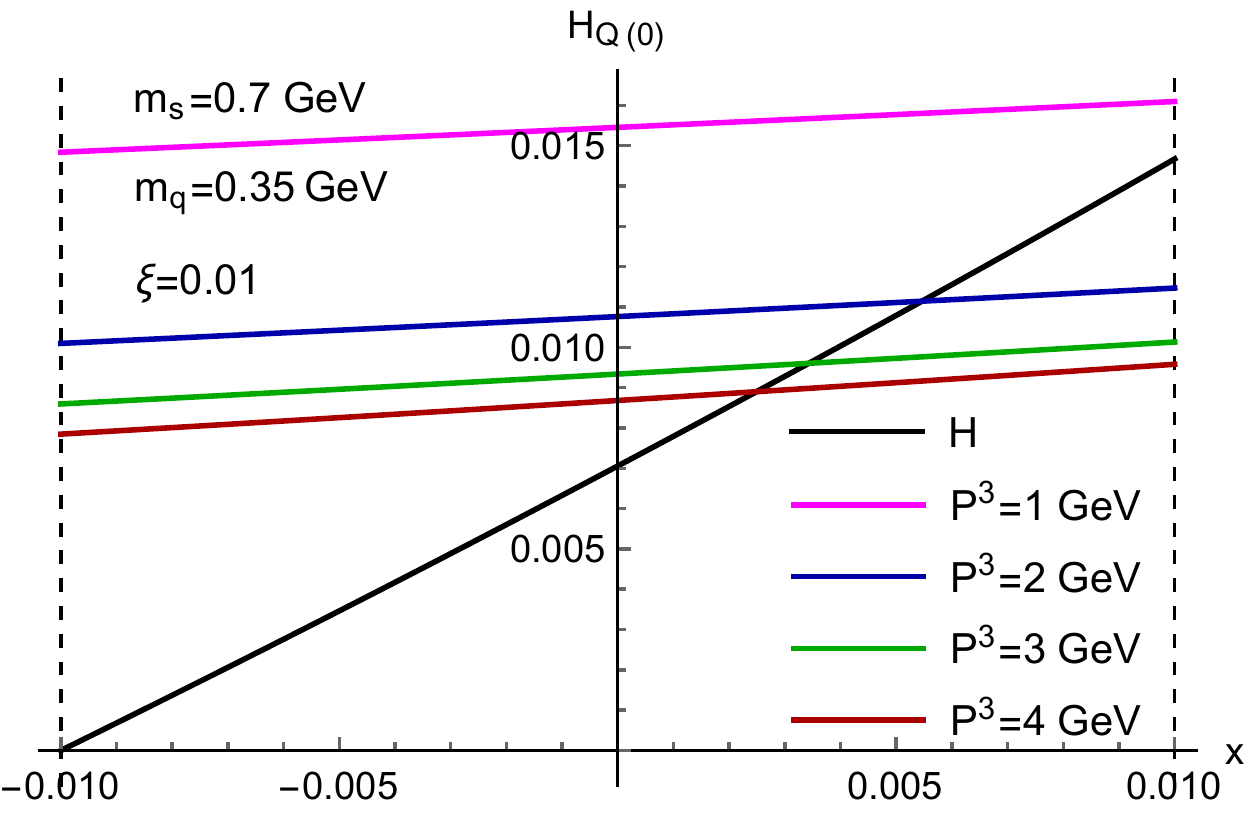}
\hspace{1.5cm}
\includegraphics[width=6.5cm]{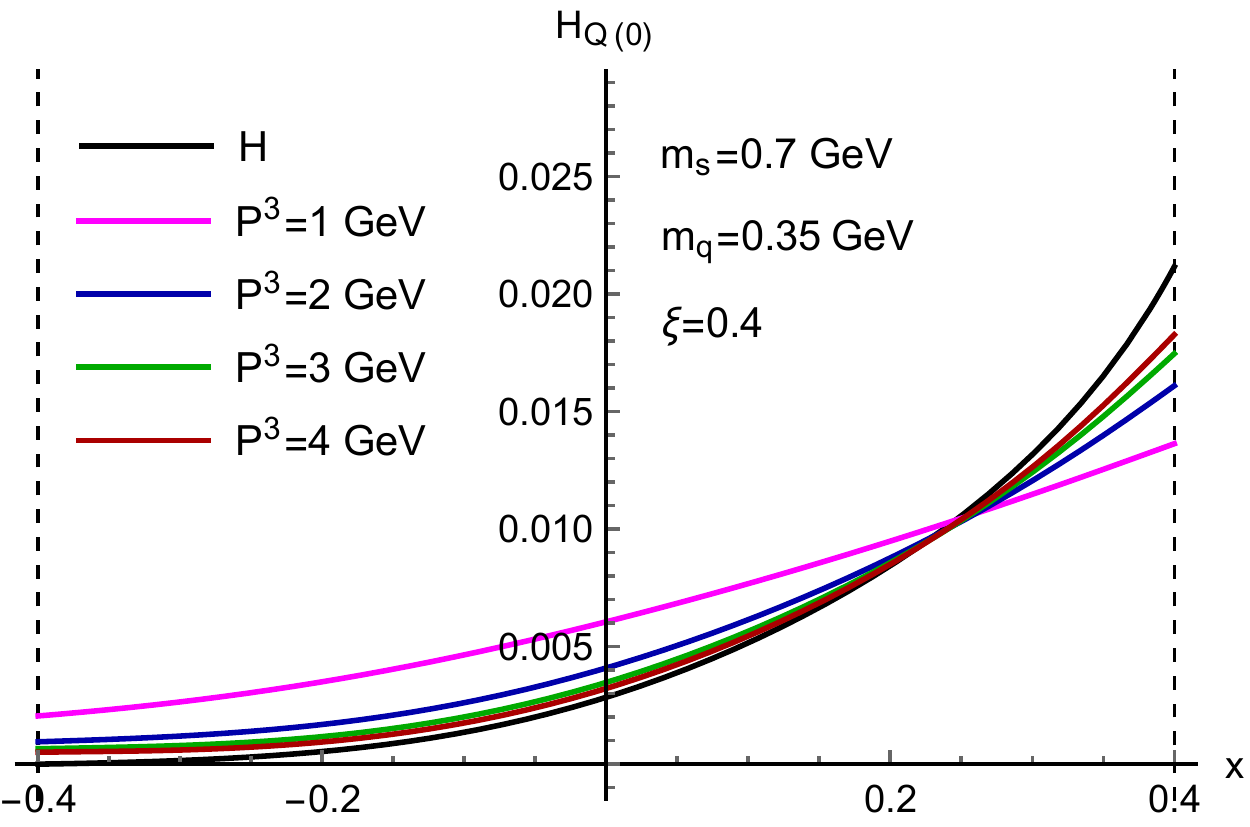}
\end{center}
\caption{Quasi-GPD $H_{{\rm Q}(0)}$ in the ERBL region for different values of $P^3$. 
Left panel: results for $\xi = 0.01$.
Right panel: results for $\xi = 0.4$.
The standard GPD $H$ is shown for comparison.}
\label{f:HQ_ERBL}
\end{figure}

\begin{figure}[t]
\begin{center}
\includegraphics[width=6.5cm]{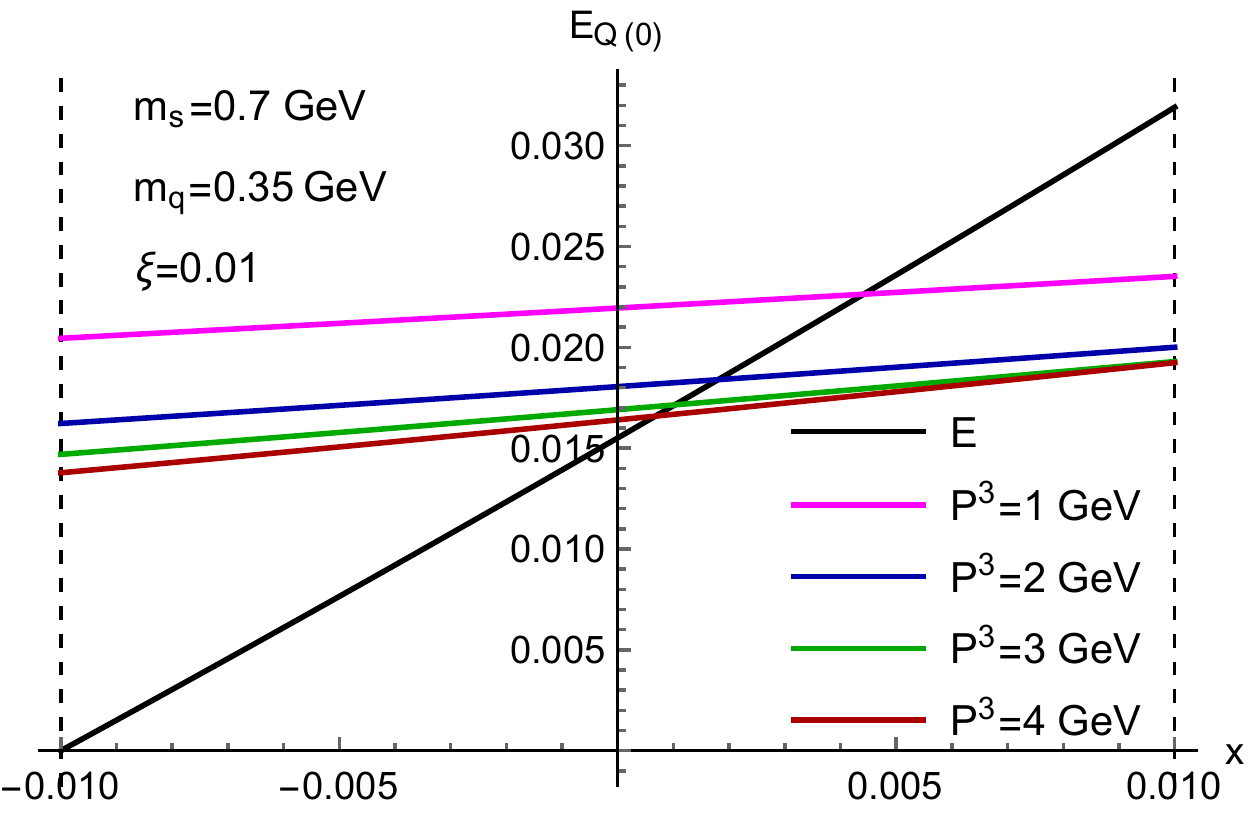}
\hspace{1.5cm}
\includegraphics[width=6.5cm]{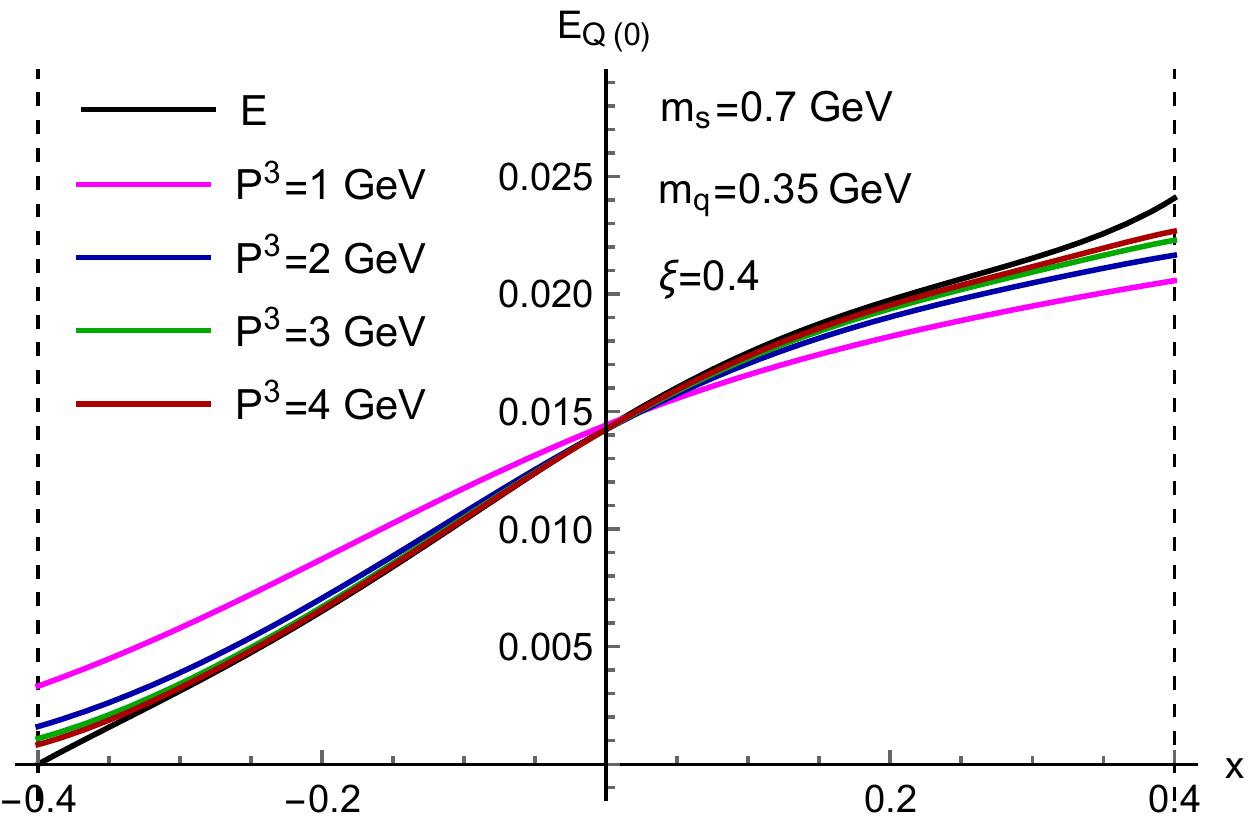}
\end{center}
\caption{Quasi-GPD $E_{{\rm Q}(0)}$ in the ERBL region for different values of $P^3$. 
Left panel: results for $\xi = 0.01$.
Right panel: results for $\xi = 0.4$.
The standard GPD $E$ is shown for comparison.}
\label{f:EQ_ERBL}
\end{figure}

Results of the quasi-GPDs for just the ERBL region are shown in  Fig.~\ref{f:HQ_ERBL} and Fig~\ref{f:EQ_ERBL}. 
(The plots are for the definitions with $\gamma^0$, but the general conclusions apply also to the case of $\gamma^3$.)
If $\xi$ is very small, there are significant differences between the quasi-GPDs and the standard GPDs for the entire ERBL region, with the relative differences becoming largest as $x$ approaches $- \, \xi$.
This problem is the GPD counterpart of the problem for quasi-PDFs around $x = 0$.
For very small $\xi$, in the SDM the standard GPDs (rapidly) go to zero at $x = - \, \xi$ within a very narrow $x$-range.
The quasi-GPDs in contrast are much smoother in that range, even for the largest $P^3$ value shown in the plots.
On the other hand, for large $\xi$ the quasi-GPDs converge very well to the standard GPDs for a large part of the ERBL region, which can be considered an encouraging result.

\section{Summary} 
\label{s:summary}
In this paper we have calculated quasi-GPDs in the SDM.
Like quasi-PDFs, which have recently been explored very intensely, quasi-GPDs are defined through spatial correlation functions that can be computed in lattice QCD.
We have focused on the twist-2 vector GPDs $H$ and $E$ and leave as future work the study of the remaining leading-twist GPDs. 
In each case we have considered two definitions for the quasi-GPDs by using the matrix $\gamma^0$ or $\gamma^3$ in the underlying quark-quark correlator.
In the forward limit one obtains the quasi-PDFs $f_{1,{\rm Q}(0/3)}$ as a byproduct. 
For all quasi-GPDs we have recovered the analytical results of the corresponding standard GPDs in the limit $P^3 \to \infty$.
This outcome further supports the idea of using quasi-GPDs to get information on standard GPDs in lattice QCD.
All results for quasi distributions are continuous, and we have argued that, in the SDM and similar approaches, this feature should persist at higher twist.
We have also found that the cut-diagram approach, which has frequently been used to compute parton distributions in spectator models, must be taken with care in the case of quasi-distributions.
For $P^3 \gtrsim 2 \, \textrm{GeV}$ the numerical results for the quasi distributions defined with $\gamma^0$ and with $\gamma^3$ are very similar.
For a wide $x$-region the quasi-GPDs are reasonably close to the standard GPDs.
This includes the ERBL region, provided that the skewness variable is not too small.
On the other hand, like for the unpolarized quasi-PDFs, at large $x$ the differences are significant for both $H_{\rm Q}$ and $E_{\rm Q}$, with the discrepancy increasing as $\xi$ gets larger.
We have also verified that the main conclusions based on the numerics are not affected if the free parameters are varied within reasonable limits.
In general, we believe it is worthwhile to further explore quasi-GPDs from a conceptual point of view as well as numerically in lattice QCD and in models.

\begin{acknowledgments}
This work has been supported by the National Science Foundation under grant number PHY-1516088.
The work of A.M.~has also been supported by the U.S. Department of Energy, Office of Science, Office of Nuclear Physics, within the framework of the TMD Topical Collaboration.
\end{acknowledgments}

\appendix


\end{document}